\documentclass[11pt,a4paper]{article}
\usepackage[margin=1in]{geometry}

\usepackage{amssymb}
\usepackage{selinput}
\usepackage{url}
\usepackage{amsmath}
\usepackage{ifthen}
\usepackage{color}
\usepackage{graphicx}
\usepackage{multirow}
\usepackage{authblk}
\usepackage{hyperref}
\usepackage{natbib}

\DeclareRobustCommand{\old}[1]{}
\DeclareRobustCommand{\new}[1]{#1}

\begin{document}

\title{Physics-based deep learning for non-stationary adaptive multiple suppression}

\renewcommand{\thefootnote}{\fnsymbol{footnote}}

\author[1]{Chao Li}
\author[1]{Yangkang Chen}
\affil[1]{The University of Texas at Austin}


\maketitle
\begin{abstract}
Surface-related multiple elimination (SRME) is an indispensable processing step in marine seismic exploration, comprising two main components: multiple prediction and adaptive subtraction. While the multiple prediction is theoretically data-driven and relies on the data itself, the adaptive subtraction step requires a fairly subjective manual control. Our approach introduces a purely data-driven and physics-based deep learning (PDL) algorithm for improving the adaptive multiple subtraction to significantly enhanced performance. There are two main innovations in the proposed deep-learning (DL) multiple subtraction approaches: first, being data-driven and self-supervised, which removes the dependence on inflexible label preparation and model training processes; secondly, being governed by the adaptive matching filtering (AMF) equation to guide the training process. The DL module serves as a strong regression filter analogous to the traditionally known AMF. We segment a moderately sized seismic profile and an initial multiple model into a large number of patches using a patching scheme for self-contained training. The number of patches is controlled by the patch size and the shift size to enable a sufficient number of training samples required by mainstream DL platforms. The patching-based self-training and regression approach, PDL, is inherently a non-stationary matching approach that aims to obtain a globally optimal matching from the initial multiple models to the input full wavefield data through iterative patching-based training. It is also closely connected with the non-stationary regression approach, which was developed almost two decades ago. However, the PDL approach achieves even stronger regression performance due to the backpropagation-based optimization mechanism in DL, and has shown significantly improved multiple subtraction performance over a number of synthetic and real data examples.
\end{abstract}

\section{Introduction}
Seismic multiples are unwanted seismic reflections that undergo more than one reflection across the subsurface formation interfaces, which can severely contaminate the primary seismic reflections and cause spurious structures on the seismic images \cite{weglein2005multiple}. Removal or suppression of these seismic multiples is particularly significant in marine seismic exploration; therefore, numerous studies have been conducted in recent decades to mitigate the influence of seismic multiples \cite{lin20043d,baumstein2005practical,lopez20153d,qu2021training,licmul2024}. There are two main types of multiples: 1) surface-related multiples, meaning that the reflections travel between the sea surface and subsurface layers (e.g., the water-bottom multiples going through multiple reflections between surface and sea floor) \cite{wiggins1988attenuation,spadavecchia2013water,zhang2020reverse}; 2) interbed multiples, indicating the reflections that bounce between layers within the subsurface (e.g., between two high-impedance layers) \cite{verschuur1996removal,jakubowicz1998wave,griffiths2011applications,sun2016interbed,zhang2021local}.  

One of the simplest ways to attenuate multiples is the Radon transform approach. \cite{hampson1986inverse} showed the possibility of applying the parabolic Radon transform to eliminate the multiples. The principle of the parabolic Radon transform is that the primary reflections can be easily flattened after normal moveout (NMO) correction, while the long-period multiples cannot be flattened with residual moveout that can be modeled by the parabolic Radon transform. Another benefit of the parabolic Radon transform is its computational advantages when implemented in the frequency domain. \cite{foster1992suppression} applied the hyperbolic Radon transform to eliminate multiple reflections due to its better distinction between primary and multiple reflections despite its high computational cost. To boost the resolution in the Radon domain for a better separation, a number of studies based on the $L_1$-norm boosted high-resolution Radon transform were developed over the years \cite{sacchi1999fast,herrmann2000aliased,moore2002stable,hugonnet20093d,chen2021fast}. \cite{ma2020multiple} developed the high-order high-resolution Radon transform for better protecting the amplitude fidelity when suppressing the multiples in the transformed domain. To further explore the subtle difference between primary and multiple reflection in the time, space, and slope dimensions, \cite{ghosh2012multiple} developed the $t-x-p$ domain approach for their separation by conducting a non-stationary regression in each frequency slice to decompose the space-dependent spectra into the space-slope domain. The Radon transform methods, however, are limited to deep-water environments where the primary and multiple reflections are distinctly separated in the velocity spectrum. When multiples overlap heavily with the primary reflections, e.g., the shallow water case, the closed-loop surface-related multiple elimination (CL-SRME) method plays a critical role \cite{lopezg2015,zhangd2021cl}.

The SRME method was proposed in \cite{verschuur1992adaptive} for addressing the challenge in suppressing surface-related multiples, particularly in marine seismic exploration. It is grounded in the principle that surface-related multiples can be predicted using only the recorded seismic data by autoconvolution \cite{anstey1966sectional,anstey1966part}, without requiring subsurface velocity or structural models. In SRME, the initial estimate of multiples is predicted by convolving recorded seismic traces with themselves. However, these predicted multiples may exhibit local time shifts and substantial amplitude mismatch that require a follow-up matching filter to make the multiples optimally fit the observed raw seismic data. The matching filter adjusted multiple models, normally using a least-squares method, is directly removed from the raw data, achieving the goal of multiple suppression. The SRME has been a state-of-the-art approach for marine seismic data processing for nearly three decades. Despite its high effectiveness, there are still challenges when applying it in practice. In reality, one problem is the strong non-stationarity of seismic data, which invalidates the use of a fixed-length stationary matching filter for adaptive multiple suppression. As an improvement, \cite{fomel2009adaptive} proposed a non-stationary matching filter to handle the inherent non-stationarity of seismic data, by formulating the matching filter problem as a regularized non-stationary regression problem, where the high-dimensional non-stationary regression coefficients are solved via an iterative shaping regularization via a smoothness constraint.

In this work, we focus on the suppression of the surface-related multiples based on the SRME framework. We follow the data-driven autoconvolution-based prediction procedure \cite{fomel2009adaptive} to estimate the initial multiples, and then apply a matching filter to remove the multiples from the raw seismic records adaptively. We propose a physics-based deep learning (PDL) algorithm to replace the commonly used stationary least-squares matching filter or the non-stationary regression filter, achieving better matched subtraction performance with improved preservation of primary reflection signals and cleaner removal of multiples. The PDL method is a pure self-supervised learning approach and thus can be readily applied to any real seismic datasets. \old{The explicit adaptive matching filtering formulation guides the PDL and provides the theoretical foundation of the learned mapping relationship between the input (e.g., estimated multiple model) and target data samples (e.g., observed data)}\new{The explicit adaptive matching filtering formulation guides the training process and provides the physical basis for learning the nonlinear mapping from the input data (e.g., the predicted multiple model) to the observed seismic data.}\old{thus explaining the guaranteed convergence and stability of the training.} \new{Furthermore, }the patching scheme is used to provide sufficient training samples for learning a stable, nonstationary, and globally optimal matching filter to suppress the multiples without damaging primary energy. We introduce in detail the principle, inherent physics, and implementation workflow of the proposed PDL in the Methodology section. A number of representative synthetic examples with increasing complexity and real data examples are used to demonstrate the efficacy of the proposed method and benchmark it against the state-of-the-art non-stationary matching filtering method. We also investigate in detail how various hyperparameters affect the final multiple suppression performance. Because of the physics-based nature, common deep-learning-related hyperparameters have limited influence on the matching performance, but the patching-related hyperparameters indeed have some effect on the accuracy of the learned matching filter, such as the patch size in both time and space dimensions and the shift size. We also investigate the influence of noise levels on the multiple suppression effect in our study.

\section{Methodology} 
\subsection{Matching filter}
Suppose we have an estimated multiple model, $\mathbf{s}$, e.g., from SRME,
we intend to design a matching filter $\mathbf{m}$ \cite{fomel2009adaptive} that is applied onto $\mathbf{s}$, so that the adjusted signal $\mathbf{m}*\mathbf{s}$ best resembles the target, i.e., the observed data $\mathbf{d}$. This process corresponds to a stationary regression problem
\begin{equation}
\label{eq:sta}
d(n) = \sum_{k=1}^{K} s(n-k+1)m(k), \quad n=1,2,\cdots,N
\end{equation}
where $n$ is the index of multiple model $\mathbf{s}$ or target data $\mathbf{d}$,  $N$ is the number of samples, and $K$ is the length of the matching filter.  In a matrix vector form, this matching filter process corresponds to the following problem
\begin{equation}
\label{eq:inv}
\mathbf{d} = \mathbf{S}\mathbf{m},
\end{equation}
where $\mathbf{S}\in\Omega^{(2N-1)\times N}$ is the convolution matrix composed from vector $\mathbf{s}$:
\begin{equation}
\mathbf{s}=
\left[\begin{array}{cccc}
s(1) & 0 & \cdots & 0 \\
s(2) & s(1) & \cdots & 0 \\
s(3) & s(2) & \cdots & 0 \\
\vdots & \vdots& \cdots & \vdots \\
s(N) & s(N-1) & \cdots & 0 \\
0 & s(N) & \cdots & 0 \\
\vdots & \vdots& \cdots & \vdots \\
0 & 0 & \cdots & s(N) \\
\end{array}\right].
\end{equation}
Under the least-squares inversion framework, equation \ref{eq:inv} can be solved as \cite{fomel2009adaptive} 
\begin{equation}
\hat{\mathbf{m}} = (\mathbf{S}^T\mathbf{S})^{-1}\mathbf{S}^T\mathbf{d}.
\end{equation}
Regardless of the size or heterogeneity of the multiple/data, the matching filter normally has a fixed length and aims to correct the amplitude and time shift between the initial estimate of the multiple and multiple-contaminated observed data. Considering the stationarity of the matching filter spatially and temporally, equation \ref{eq:sta} is referred to as a stationary matching filter.

\subsection{Non-stationary matching filter}
To make the matching filter spatially and temporally non-stationary, equation \ref{eq:sta} can be slightly modified to enable the non-stationarity:
\begin{equation}
\label{eq:nsta}
d(n) = \sum_{k=1}^{K} s(k,n)m(k,n), \quad n=1,2,\cdots,N,
\end{equation}
which indicates the matching filter $m(k,n)$ is temporally/spatially variable. Here, $n$ denotes the temporal or spatial index when the multiple/data model is viewed as a long flattened vector. Solving equation \ref{eq:nsta} is significantly more challenging than solving equation \ref{eq:sta}, because the model $m(k,n)$ contains $KN$ unknown parameters, whereas only $N$ observations are available in $d(n)$. To solve equation \ref{eq:nsta}, one needs to first transform it into a matrix-vector form as
\begin{equation}
\label{eq:nsta2}
\mathbf{d} = \left[\begin{array}{cccc}
\mathbf{S}_1 & \mathbf{S}_2 & \cdots & \mathbf{S}_K
\end{array}\right] \left[ \begin{array}{c}
\mathbf{m}_1\\
\mathbf{m}_2\\
\vdots\\
\mathbf{m}_K
\end{array}\right], 
\end{equation}
where $\mathbf{S}_k$ denotes a diagonal matrix composed of a shifted vector $\mathbf{s}$ with a shift lag of $k$, $k=1,2,\cdots,K$. $\mathbf{m}_k$ denotes the non-stationary matching filter of $k$th component. 
\cite{fomel2009adaptive} used the shaping regularization algorithm \cite{fomel2007shape} to solve this highly under-determined inverse problem via the following approach
\begin{equation}
\label{eq:iter}
\tilde{\mathbf{m}}_{i+1} = \mathbf{T}(\tilde{\mathbf{m}}_{i} + \tilde{\mathbf{S}}^T (\mathbf{d} - \tilde{\mathbf{S}}\tilde{\mathbf{m}}_{i+1}) ),
\end{equation}
where $i$ denotes the iteration number and $\tilde{\mathbf{S}}$ is the block forward operator ($\tilde{\mathbf{S}}=[\mathbf{S}_1, \mathbf{S}_2, \cdots, \mathbf{S}_K]$) and $\tilde{\mathbf{m}}$ is the block matching filter vector ($\tilde{\mathbf{m}}=[\mathbf{m}_1, \mathbf{m}_2, \cdots, \mathbf{m}_K]^T$). $\mathbf{T}$ is the shaping operator \cite{fomel2007shape}, which can impose additional constraints into the estimated model and it is normally chosen as a multi-dimensional triangle smoothing operator \cite{orthon2021}.

\subsection{Physics-based deep learning: Neural-network matching filter}
The physics behind the non-stationary matching filter (equation \ref{eq:nsta2}) is that a matching filter with spatially and temporally variable filter coefficients is solvable for correcting the phase shift and amplitude errors in the multiple subtraction problem, as demonstrated in \cite{fomel2009adaptive}. The absolute value of the non-stationary matching filter coefficients indicates the level of misfit between the two signals to be matched, while the change in the coefficients indicates the non-stationarity level. Traditional deterministic non-stationary matching filter assumes a smooth non-stationary change, as guaranteed by the triangle-smoothing shaping operator ($\mathbf{T}$ in equation \ref{eq:iter}), which works well in a wide range of problems and datasets. However, enforcing the smoothness of the non-stationary matching filter coefficients limits the matching performance in highly non-stationary scenarios, especially in capturing fast local changes. Here, we enhance its capability by proposing to use a stronger regression algorithm to bypass the local smoothness constraint as used in earlier studies. First, we can reformulate equation \ref{eq:nsta2} in a slightly different way:
\begin{equation}
\mathbf{d} = \mathcal{M}(\mathbf{s}, \theta),
\label{eq:mstheta}
\end{equation}
where $\mathcal{M}$ represents a mapping operator between $\mathbf{s}$ and $\mathbf{d}$ and $\theta$ denotes the estimable parameters (e.g., weight matrices and bias vectors). Equation~\ref{eq:mstheta} indicates the non-stationary matching is achieved by applying a matching-based operator to the multiple models $\mathbf{s}$, and we know from earlier analysis that $\mathcal{M}$ is solvable using deterministic methods. Considering the strong ability of deep learning on nonstationary and nonlinear data matching, we propose a neural network (NN) based matching filter to substitute the regularized non-stationary matching $m(k,n)$:
\begin{equation}
\label{eq:nn}
d(n) = \text{NN}(\theta, s(n)), \quad n=1,2,\cdots,N,
\end{equation}
where $\text{NN}$ is a deep neural network and specifically $\theta$ is the trainable neurons. The $\text{NN}$ operator can be quite flexible, considering many of the latest advances in deep learning architecture for capturing highly complicated mapping relationships. One of the simplest examples is the multi-layer feed-forward architecture, which has a basic formulation as
\begin{equation}
\label{eq:nn2}
\text{NN}(\theta, \mathbf{s} =  \sigma_l[\mathbf{W}_l(\cdots \sigma_1[\mathbf{W}_1(\mathbf{s})+\mathbf{b}_1])+\mathbf{b}_l]), \quad l=1,2,\cdots,L,
\end{equation}
where $L$ denotes the number of layers, $\sigma_l$ denotes the activation function of the $l$th layer, $\mathbf{W}_l$ and $\mathbf{b}_l$ are the weight matrix and bias vector of the $l$th layer. 

To ensure effective training, we use the patching scheme as described by previous research \cite{omar2021gp1,omar2023,saado2024} (see an illustration in Figure~\ref{fig:pdlsketch}) to prepare a large number of patches, with the number of patches controlled by the patch size and shift size. The patching scheme helps reduce computational cost during network training, and by processing each patch independently, the network learns to match inputs and outputs in a localized manner. This encourages the model to focus more effectively on local feature extraction, thereby improving the overall prediction accuracy. The non-stationary matching filter within each patch is learned over the tens of thousands of training samples, thereby obtaining a globally optimal mapping across all the training patches. Despite the flexibility in designing the neural network, e.g. number of layers, activation functions, batch size, and other training parameters, the proposed PDL method is guaranteed to converge and be stable because of the non-stationary matching filtering based physics as depicted by equations \ref{eq:sta}, \ref{eq:nsta}, and \ref{eq:nsta2}. It is worth noting that the proposed PDL bypasses the assumptions of smoothly varying non-stationarity and the smooth matching filter coefficients, because the network parameters are estimated through a standard back-propagation optimization algorithm, without the smoothness constraint. However, the batch size during the training process enforces the smoothness constraint of the gradient update, but in a non-local way, thus PDL is particularly effective in capturing the highly oscillating local variations. From the stability and constraint standpoint, PDL leverages non-local smoothness constraints, analogous to those in some existing works in the denoising literature \cite{buades2011non,wanghang2020geo3}, to ensure a stable, globally optimal matching filtering. The complete workflow of the PDL method for multiple suppression is shown in Figure \ref{fig:pdlsketch}, which is composed of two steps: 1) patching to prepare training samples; 2) physics-based neural network matching filtering to suppress multiples.

\subsection{A benchmark comparison between non-stationary regression}
The stationary, non-stationary, and neural-network matching filters are all fundamentally regression algorithms. To have a clear understanding of the principles, strengths, and weaknesses of different approaches, we prepared a simple synthetic benchmark example, inspired by \cite{fomel2009adaptive}. Figure \ref{fig:comp}(a) shows 50 randomly chosen scattering points following a normalized quadratic polynomial with added random Gaussian noise. This example will be used as a line-fitting example to benchmark across all related methods. It is worth noting that, if in this regression problem, we consider the horizontal axis coordinate ($x$) as the base signal ($\mathbf{s}$) and consider the scattering points as the target signal ($\mathbf{d}$), then this regression problem is analogous to the non-stationary matching filtering problem.  Stationary line fitting obviously cannot capture the trend of the scattering points well (we omitted the plot in this case). A better way to improve the fit is to include nonlinear terms (e.g., $x^2$ in this case), as shown in Figure \ref{fig:comp}(a), or to break the points into two windows with each window characterized by a linear fitting, as shown in Figure \ref{fig:comp}(b). However, both methods cannot capture the local oscillation very well because of the random noise that is difficult to simulate using a fixed polynomial basis term (e.g., $x$, $x^2$, or $x^3$). The regularized non-stationary regression \cite{fomel2009adaptive} captures the local oscillation in the data much better, as shown in  Figure \ref{fig:comp}(c), because of the use of the non-stationary matching/regression coefficients. Figure \ref{fig:comp}(d) shows the fitted curve using the PDL matching/regression, which shows the best fitting performance, and is even better than the regularized non-stationary regression, especially in the edge points (left corner around $x=0$ and right corner around $x=1.0$). This example clearly demonstrates the capability of each individual method in terms of fitting non-stationary and noise-contaminated data.

\section{Result} 
In this section, we use two synthetic examples and a field dataset to experimentally evaluate the proposed PDL method over the classic nonstationary least-square matching filter (LSMF) \cite{fomel2009adaptive} in terms of accuracy and fidelity. The network we use contains two dense layers with ReLU and linear activation functions, respectively. The number of neurons in each layer is (Patch\_t $\times$ Patch\_x)/4 and Patch\_x $\times$ Patch\_t for the first and second layer, respectively, where Patch\_t and Patch\_x stand for the patch size of t and x coordinates.

\subsection{Synthetic data}
The first synthetic data is a shot gather after normal moveout correction (NMO) with flattened primaries and curved multiples, which is similar to the example used in \cite{fomel2009adaptive}. The synthetic data is generated by adding multiple flat primary reflections and two curved multiples with residual moveout. Figure~\ref{fig:syndata1}a shows the real primary reflections, which are the ground truth. Figures~\ref{fig:syndata1}b and~\ref{fig:syndata1}c show the original data and the estimated multiples with amplitude and phase differences. Then we apply the classic LSMF and PDL methods to the synthetic data shown in Figure~\ref{fig:syndata1} and the corresponding demultipled results are depicted in Figure~\ref{fig:synpre1}. The data size is 751 $\times$ 121 (e.g., Time $\times$ Trace) and we set the patch size and shift size to 10 and 1 for the first synthetic example. Figures~\ref{fig:synpre1}a-\ref{fig:synpre1}d display the estimated primary reflections and the removed multiples. Notably, we can see the energy leakage from the multiples in Figure~\ref{fig:synpre1}a (e.g., marked by the red arrows). To quantify the performance of classic LSMF and PDL, as shown in Figure~\ref{fig:synpre1}, we calculate the Mean Squared Error (MSE) and the Structural Similarity Index Measure (SSIM)  between the final estimated primaries and the ground truth. Comparing MSE and SSIM, we can recognize that the proposed PDL method outperforms the classic LSMF in terms of accuracy and structural similarity. In addition, we then extract and magnify the area marked by the blue dotted rectangle. Figure~\ref{fig:zoomsyn1} shows the zoomed sections for further comparison, with the obvious improvements already marked. Compared to Figure~\ref{fig:zoomsyn1}d, there are more visible multiple residuals in Figure~\ref{fig:zoomsyn1}c. More importantly, from the perspective of amplitude fidelity, the PDL method causes fewer errors for effective signals (e.g., the flat event marked by red arrows), indicating that PDL can establish a non-linear mapping relationship, enabling it to handle the complexity and non-stationarity of seismic data. Figure~\ref{fig:synspe1} shows the spectra comparison of the first synthetic data, where we can see that, except for some multiple related energy attenuation, the spectra of primaries estimated by LSMF and PDL show a similar trend, demonstrating that the proposed method does not distort the data in the frequency domain.

\par Next, we use a more complex Pluto dataset \cite{fomel2009adaptive} in common offset domain with an temporal sampling of 8 ms to validate the effectiveness of the proposed PDL method. Figure~\ref{fig:syndata2} shows the original seismic data with multiples and the multiple model estimated using SRME method, where we can see obvious amplitude and phase differences between the real multiple shown in Figure~\ref{fig:syndata2}a and the estimated multiple model shown in Figure~\ref{fig:syndata2}b. The data size is 1126 $\times$ 818 (e.g., Time $\times$ Trace) and we set the patch size and shift size to 32 and 8 for the second synthetic example. Figure~\ref{fig:synpre2} displays the final predicted primaries and the removed multiples. Figures~\ref{fig:synpre2}a and~\ref{fig:synpre2}b depict the primaries and we can pay attention to the areas marked by the blue dotted rectangles, where the classic LSMF method results in unexpected primary damage, leading to inferior fidelity during multiple suppression. Figures~\ref{fig:synpre2}c and~\ref{fig:synpre2}d depict the removed multiples corresponding to the Pluto dataset. Notably, compared to the PDL method, we can see that the LSMF method causes some visible and strong primary reflections energy leakage (marked by red arrows). To provide a more intuitive comparison, Figure~\ref{fig:zoomsyn2} exhibits the magnified section of the original data and the estimated primaries obtained by the LSMF method and the PDL method, respectively. Compared to the original data, we can easily recognize that the proposed PDL method has a superior ability to preserve waveform continuity (e.g., marked by the red arrows). Moreover, it is worth noting that the proposed PDL method has a stronger capability to preserve weak and effective signals (e.g., marked by the blue arrows), which causes fewer errors for further seismic processing routines. Figure~\ref{fig:synspe2} shows the spectra comparison of raw data and the estimated primaries using LSMF method and PDL method, respectively. The raw data (blue curve) exhibits significant energy around 15–30 Hz, which corresponds to the strong multiple reflections visible in Figure~\ref{fig:syndata2}a. Both LSMF (red curve) and PDL (black curve) reduce this multiple-dominated mid-frequency band, consistent with the removal of the strong surface-related and interbed multiples shown in Figures~\ref{fig:synpre2}c and~\ref{fig:synpre2}d. Compared with LSMF, the PDL output aligns even more closely with the raw data in the high-frequency portion, demonstrating that the neural network does not excessively attenuate or amplify high-frequency primary energy

\subsection{Field data}
In this section, we apply the proposed PDL to a field data example from the Viking Graben \cite{madiba2003} for further assessment. The data size is 1500 $\times$ 120 $\times$ 1012 (e.g., Time $\times$ Offset $\times$ Shot) with a temporal sampling rate of 4 ms and we set the patch size and shift size to 64 and 8 for the field data. Then we conduct the classic LSMF and the proposed PDL methods across offsets (e.g., common offset domain) for multiple suppression, and the final results are displayed in Figure~\ref{fig:3dshotfld1}. Figures~\ref{fig:3dshotfld1}a and \ref{fig:3dshotfld1}b show the original shot gathers and the estimated multiples using the SRME method. Figures~\ref{fig:3dshotfld1}c-\ref{fig:3dshotfld1}f display the predicted primary waves and the removed multiples via the classic LSMF method and PDL method, respectively. It is noticeable that compared to the classic LSMF method, the proposed PDL method can suppress multiples more effectively and lead to fewer multiple residuals. \new{To further illustrate the improvement achieved by the proposed PDL method, Figure~\ref{fig:fldstacks} compares the stacked sections over the 0–4 s time interval, which includes the main target reservoir (1.3–2.5 s) \cite{madiba2003}, obtained from the different datasets. As indicated by the red arrows, the stacked section of the raw data is severely contaminated by multiple reflections, while the LSMF result still contains noticeable residual multiple energy, leading to image degradation. In comparison, the stacked section produced by the proposed PDL method (e.g., Figure~\ref{fig:fldstacks}c) exhibits fewer residual multiples and improved primary continuity, resulting in a cleaner image of the target interval. }

\par \new{To further provide an intuitive demonstration, we first extract a single-shot gather as shown in Figure~\ref{fig:shotfld1} for further comparison}.  Figures~\ref{fig:shotfld1}a and~\ref{fig:shotfld1}b depict the original data and the estimated multiples in the common shot domain. Figures~\ref{fig:shotfld1}c and~\ref{fig:shotfld1}d show the results after multiple separation. Note the events highlighted by the red arrows, where we can clearly see that the multiples shown in Figure~\ref{fig:shotfld1}d are nearly eliminated, indicating the superiority of the proposed PDL method. Notably, although some channel-to-channel energy inconsistencies can be observed in Figure~\ref{fig:shotfld1}, the multiple removal is implemented in the common-offset domain, where such channel-related energy variations are less evident. Therefore, their influence on the demultiple results is relatively limited in our current implementation. Figure~\ref{fig:fldprespe1} presents a spectral comparison among the raw shot gather (Figure~\ref{fig:shotfld1}a), the primary estimate obtained using the classic LSMF method (Figure~\ref{fig:shotfld1}c), and the primary estimate produced by the proposed PDL method (Figure~\ref{fig:shotfld1}d). The raw data spectrum (blue curve) exhibits strong multiple-related energy, particularly in the 10–35 Hz band. After demultiple, both LSMF and PDL reduce the amplitude in this frequency range, which is expected since surface-related multiples dominate these mid-frequency components. Importantly, the PDL spectrum (black curve) closely follows the LSMF spectrum (red curve) across all frequencies, indicating that the network does not introduce artificial attenuation or amplification in any specific band. Overall, the spectral comparison demonstrates that the proposed PDL method suppresses multiple-related spectral energy while preserving the broadband characteristics of the primaries, without disproportionately attenuating any frequency band.

\par Additionally, Figure~\ref{fig:flddata1} shows one common offset gather extracted from the field data with multiples and the estimated multiple through the SRME method, and the final predicted results are shown in Figure~\ref{fig:fldpre1}. In particular, Figures~\ref{fig:fldpre1}a and~\ref{fig:fldpre1}b show the predicted primaries and Figures~\ref{fig:fldpre1}c and~\ref{fig:fldpre1}d correspond to the removed multiples. Comparing Figures~\ref{fig:fldpre1}a and~\ref{fig:fldpre1}b, we see that the classic LSMF method exhibits inferior performance, and the blue dotted rectangle marks multiple leakages (e.g., visible residual multiples). In addition, we can also pay attention to the marked area in Figures~\ref{fig:fldpre1}c and~\ref{fig:fldpre1}d, where we can see that the removed multiples retain better continuity and consistency of the amplitude. To make a further comparison, Figure~\ref{fig:fldpresim1} shows local similarity coefficients \cite{sergeylocal2007} between the original data and the estimated multiples using the LSMF method and the PDL method, respectively. As expected, the removed energy exhibits high similarity along the major multiple trends and low similarity between the removed energy and primary-dominated regions, indicating that the subtraction mainly targets multiple-related events and primary energy is preserved. To provide a further comparison, we extract part of data from Figures~\ref{fig:flddata1}a, ~\ref{fig:fldpre1}a, and~\ref{fig:fldpre1}b, and the magnified sections are shown in Figure~\ref{fig:zoomfld1}. Figure~\ref{fig:zoomfld1}a shows the original data where we can see multiple events (e.g., marked by red arrows). Figures~\ref{fig:zoomfld1}b and~\ref{fig:zoomfld1}c show the estimated primaries using classic LSMF and PDL, respectively. Notably, we can clearly see that the result obtained by the PDL method shows fewer multiple residuals, indicating that PDL is more effective than classic LSMF in dealing with the complexity and non-stationarity of the field data and leading to superior fidelity and accuracy.

\section{Discussion}
In this section, we discuss several key factors related to PDL and offer suggestions for parameter selection to enhance performance. All the experiments are based on the synthetic data shown in Figure~\ref{fig:syndata1}.
\subsection{The influence of patch size and shift size}
The proposed PDL method incorporates a patch based scheme for local feature extraction. The selection of patch size can have considerable effects on the final results. To figure out the influence of patch size, we select a patch size from a predefined range between 5-30 with an increment of 5. Next, we conduct the experiment using varying patch sizes and keep the other parameters unchanged (e.g., learning rate), and Figure~\ref{fig:patchsize}a shows the calculated MSE values. In particular, we find that the final MSE will increase with the increasing patch size. However, the rate of change of MSE with respect to patch size along different axes is different. Figure~\ref{fig:patchsize}b shows two MSE variation curves when we fix Patch\_t and Patch\_x to 5, respectively, where we can see that Patch\_x is more sensitive to MSE when Patch\_t is fixed. The synthetic data shown in Figure~\ref{fig:syndata1} contains 121 samples in the x coordinate. Therefore, Patch\_x=30 covers nearly 25$\%$ of the data along the x-axis and makes it less effective in extracting local feature maps. Based on our experiments, we recommend that the patch size of a specific axis should be smaller than 15$\%$ of the total number of sampling points. Moreover, a smaller patch size can lead to more number of training patches, increasing the computational cost. Then we fix the patch size to 10 for both axes and change the shift size from 1 to 8 to evaluate its influence. Figure~\ref{fig:shiftsize} displays the final MSE and the corresponding computational cost. Larger shift size can lead to lower accuracy but higher computational efficiency. In normal cases, we suggest conducting a parameter sensitivity analysis before data processing to determine a reliable patch size and shift size for each axis, thereby achieving the highest accuracy and efficiency.

\subsection{The influence of noise level}
Noise level is another factor that influences the final results. To intuitively quantify the noise level, we define the signal-to-noise ratio (SNR) with the following expression
\begin{equation}
\mathrm{SNR}=20\log_{10}\left(\left \| \mathbf{s}_{\mathrm{clean}}\right \|_{2}/\left \| \mathbf{s}_{\mathrm{noise}}\right \|_{2}\right),
\label{eq:snr}
\end{equation}
where $\mathbf{s}_{\mathrm{clean}}$ represents the clean seismic data and $\mathbf{s}_{\mathrm{noise}}$ is noise. We set the patch size and shift size to 10 and 1, respectively. Figure~\ref{fig:snrs} shows the noisy data with different noise levels and the corresponding prediction results. Due to the effects of noise, we can observe that some weak signals are overwhelmed, resulting in a risk of effective information loss. Figures~\ref{fig:snrs}c and~\ref{fig:snrs}d show the final predicted primaries along with the respective MSE and SSIM. It is obvious that although the final performance of PDL will degrade with the increasing noise level, the PDL method is still effective in suppressing multiples. Moreover, we think the proposed PDL method has limited ability to suppress ambient noise, and we suggest pre-processing the input seismic data (e.g., denoising) before multiple suppression to reach improved performance in real applications. We will continue our research on how to suppress multiples and ambient noise simultaneously using the PDL method in the future.

\subsection{Connection and difference with other state-of-the-art demultiple frameworks }
\new{Curvelet domain multiple adaptive subtraction is another alternative for effective multiple suppression \cite{Herrmann2008,neelamani2010}. Specifically, both curvelet-domain adaptive subtraction and the proposed PDL method aim to compensate for discrepancies between the SRME-predicted multiple model and the multiples contained in the recorded data. However, they differ in how the adaptive matching operator configured. Curvelet-domain adaptive subtraction performs the matching process in the curvelet domain in a multiscale but linear manner, where localized matching parameters are estimated for different scales, orientations, and spatial locations. Then the corrected coefficients are subsequently transformed back to the data domain through the inverse curvelet transform to recover the primary wavefield. Consequently, the effectiveness of this approach relies on the sparse multiscale representation of seismic events provided by the curvelet frame. In contrast, PDL method does not rely on a predefined sparse transform. Instead, it implicitly learns a nonlinear and nonstationary regression operator directly through patch-based self-supervised training guided by AMF. Benefiting from the ability of deep learning to learn complex nonlinear and nonstationary relationships, the proposed PDL framework may offer greater flexibility in handling discrepancies between the predicted and recorded multiples.}

\new{In addition to sparse transform-based approaches, local primary-and-multiple orthogonalization (LPMO) has also been proposed as an effective alternative for multiple suppression \cite{zhangd2019,zhangd2021f}. Commonly, LPMO follows a two-step strategy. The first stage generates an initial primary estimate using conservative SRME, where residual multiples remain in the estimated primary to avoid excessive suppression of the primary events. The residual multiple energy is subsequently estimated and further removed from the initial primary estimate through LPMO. Consequently, LPMO can be regarded as a complementary post-processing technique to conventional adaptive multiple subtraction, improving the fidelity of the recovered primary wavefield. Unlike LPMO, which relies on a two-stage workflow involving post-processing refinement, the proposed PDL framework performs adaptive multiple subtraction in a single stage. It aims to learn a more accurate nonlinear matching filter to compensate for the discrepancies between the predicted and recorded multiples, thereby enabling direct subtraction without requiring an additional post-processing step.}

\subsection{Strength and limitation of the proposed method}
In this paper, we do not use a physics-informed constraint, e.g., as in the loss function, but instead, we seek the physics behind the implicit neural network matching, i.e., $d(n) = \sum_{k=1}^{K} s(k,n)m(k,n), \quad n=1,2,\cdots,N$. We analyze the fundamental physics-based mechanism that makes this neural network matching filter highly stable and effective, implicitly replacing the conventional non-stationary matching filter. Although the proposed PDL method is mainly designed for free surface multiples, we think that it is also effective for interbed multiples, even if an approximated interbed multiple model can be provided. The PDL method is self-adaptive and can improve local data fit. However, the proposed PDL method can achieve improved performance in some specific domains (e.g., the common offset domain), where events are relatively flat. For some complex structures, the patching scheme cannot guarantee holistic recovery unless the shift size is small, which may increase computational cost. Given this point, we consider developing an advanced scheme to ensure both efficiency and accuracy for large-scale 3D applications.

\section{Conclusion}
The traditional matching filter can be ineffective in SRME for highly non-stationary seismic data, and the non-stationary matching filter can make the matching more adaptive to the spatial heterogeneity. The non-stationary matched filter, however, is based on regularized non-stationary regression and a locally smooth constraint on the regression coefficients, which may limit its effectiveness in regions with rapid spatial variations in signal energy. We have proposed a highly flexible, effective, and physics-based self-supervised deep learning approach, PDL, for addressing the non-stationary matching problem in SRME. The PDL has a straightforward formulation as its loss function, i.e., the matched filtering misfit between the multiple samples and the noise-contaminated data samples. The fundamental principle of the PDL is to learn the globally optimal matching filter for all the small patch samples, and it can adapt to both linear and non-linear misfit between the multiples and noise-contaminated data, which theoretically outperforms the conventional matching filter and the non-stationary matched filtering. Both synthetic and real data examples show promising results of the new PDL approach in preserving the primary signals and removing the multiples. The PDL method is preferably applied to common-offset gathers due to the relatively flat structures, resulting in less aliasing and lower dip angles, which are easier to extract features from using patch-based deep learning.

\section{Data Availability Statement}
Data associated with this research are available and can be obtained by contacting the corresponding author.

\bibliographystyle{unsrt}
\bibliography{my}

@article{baumstein2005practical,
  title={A practical procedure for application of 3{D} {SRME} to conventional marine data},
  author={Baumstein, Anatoly and Hadidi, Mohamed T and Hinkley, David L and Ross, Warren S},
  journal={The Leading Edge},
  volume={24},
  number={3},
  pages={254--258},
  year={2005},
  publisher={Society of Exploration Geophysicists}
}

@Article{omar2023,
  author={Omar M. Saad and Sergey Fomel and Ray Abma and Yangkang Chen},
  title = {Unsupervised deep learning for 3{D} interpolation of highly incomplete data},
  journal={Geophysics},
  year=2023,
  volume=88,
  issue=1,
  number=1,
  pages={WA189–WA200},
  doi={10.1190/GEO2022-0232.1},
}

@article{buades2011non,
  title={Non-local means denoising},
  author={Buades, Antoni and Coll, Bartomeu and Morel, Jean-Michel},
  journal={Image processing on line},
  volume={1},
  pages={208--212},
  year={2011}
}

@Article{wanghang2020geo3,
  author={Hang Wang and Yangkang Chen},
  title = {Adaptive frequency-domain non-local means for seismic random noise attenuation},
  journal={Geophysics},
  year=2020,
  volume=86,
  issue=2,
  number=2,
  pages={V143–V152},
  doi={10.1190/geo2019-0798.1},
}

@Article{omar2021gp1,
  author={Omar M. Saad and Yangkang Chen},
  title = {A fully-unsupervised and highly-generalized deep learning approach for random noise suppression},
  journal={Geophysical Prospecting},
  year=2021,
  volume=69,
  issue=4,
  number=4,
  pages={709-726},
  doi={10.1111/1365-2478.13062},
}

@Article{fomel2007shape,
  author={Sergey Fomel},
  title={Shaping regularization in geophysical-estimation problems},
  journal={Geophysics},
  year=2007,
  Volume=72,
  issue=2,
  number=2,
  pages={R29-R36},
}

@Article{orthon2021,
  author={Yangkang Chen and Sergey Fomel},
  title = {Nonstationary local signal-and-noise orthogonalization},
  journal={Geophysics},
  year=2020,
  volume=86,
  issue=5,
  number=5,
  pages={V409–V418},
  doi={10.1190/geo2020-0151.1},
}

@article{ma2020multiple,
  title={Multiple attenuation with 3{D} high-order high-resolution parabolic Radon transform using lower frequency constraints},
  author={Ma, Jitao and Xu, Guoyang and Chen, Xiaohong and Wang, Xiaoliu and Hao, Zhenjiang},
  journal={Geophysics},
  volume={85},
  number={3},
  pages={V317--V328},
  year={2020},
  publisher={Society of Exploration Geophysicists}
}

@article{qu2021training,
  title={Training deep networks with only synthetic data: Deep-learning-based near-offset reconstruction for (closed-loop) surface-related multiple estimation on shallow-water field data},
  author={Qu, Shan and Verschuur, Eric and Zhang, Dong and Chen, Yangkang},
  journal={Geophysics},
  volume={86},
  number={3},
  pages={A39--A43},
  year={2021},
  publisher={Society of Exploration Geophysicists}
}

@article{lin20043d,
  title={3{D} {SRME} application in the {G}ulf of {M}exico},
  author={Lin, Dechun and Young, Jerry and Huang, Yan},
  journal={ASEG Extended Abstracts},
  volume={2004},
  number={1},
  pages={1--4},
  year={2004},
  publisher={CSIRO Publishing}
}

@inproceedings{lopez20153d,
  title={3{D} focal closed-loop {SRME} for shallow water},
  author={Lopez, Gabriel A and Verschuur, DJ},
  booktitle={SEG International Exposition and Annual Meeting},
  pages={SEG--2015},
  year={2015},
  organization={SEG}
}

@inproceedings{moore2002stable,
  title={Stable, efficient, high-resolution Radon transforms},
  author={Moore, I and Kostov, CLEMENT},
  booktitle={64th EAGE Conference \& Exhibition},
  pages={cp--5},
  year={2002},
  organization={European Association of Geoscientists \& Engineers}
}

@article{zhang2020reverse,
  title={Reverse time migration using water-bottom-related multiples},
  author={Zhang, Yanbao and Liu, Yike and Liu, Xuejian and Zhou, Xiaopeng},
  journal={Geophysical Prospecting},
  volume={68},
  number={2},
  pages={446--465},
  year={2020},
  publisher={European Association of Geoscientists \& Engineers}
}

@article{spadavecchia2013water,
  title={Water-bottom multiple attenuation by Kirchhoff extrapolation},
  author={Spadavecchia, Emmanuel and Lipari, Vincenzo and Bienati, Nicola and Drufuca, Giuseppe},
  journal={Geophysical Prospecting},
  volume={61},
  number={4},
  pages={725--734},
  year={2013},
  publisher={European Association of Geoscientists \& Engineers}
}

@article{wiggins1988attenuation,
  title={Attenuation of complex water-bottom multiples by wave-equation-based prediction and subtraction},
  author={Wiggins, J Wendell},
  journal={Geophysics},
  volume={53},
  number={12},
  pages={1527--1539},
  year={1988},
  publisher={Society of Exploration Geophysicists}
}

@article{zhang2021local,
  title={Local primary-and-multiple orthogonalization for leaked internal multiple crosstalk estimation and attenuation on full-wavefield migrated images},
  author={Zhang, Dong and Verschuur, DJ and Davydenko, Mikhail and Chen, Yangkang and Alfaraj, Ali M and Qu, Shan},
  journal={Geophysics},
  volume={86},
  number={1},
  pages={A7--A13},
  year={2021},
  publisher={Society of Exploration Geophysicists}
}

@article{griffiths2011applications,
  title={Applications of interbed multiple attenuation},
  author={Griffiths, MalcolM and Hembd, Jeshurun and Prigent, Herv\'{e}},
  journal={The Leading Edge},
  volume={30},
  number={8},
  pages={906--912},
  year={2011},
  publisher={Society of Exploration Geophysicists}
}

@inproceedings{verschuur1996removal,
  title={Removal of interbed multiples},
  author={Verschuur, DJ and Berkhout, AJ},
  booktitle={58th EAGE Conference and Exhibition},
  pages={cp--48},
  year={1996},
  organization={European Association of Geoscientists \& Engineers}
}

@incollection{jakubowicz1998wave,
  title={Wave equation prediction and removal of interbed multiples},
  author={Jakubowicz, Helmut},
  booktitle={SEG Technical Program Expanded Abstracts 1998},
  pages={1527--1530},
  year={1998},
  publisher={Society of Exploration Geophysicists}
}

@article{sun2016interbed,
  title={Interbed multiple prediction on land: Which technology, and which domain?},
  author={Sun, Jian and Innanen, KAH},
  journal={Recorder},
  volume={41},
  pages={24--29},
  year={2016}
}

@article{fomel2009adaptive,
  title={Adaptive multiple subtraction using regularized nonstationary regression},
  author={Fomel, Sergey},
  journal={Geophysics},
  volume={74},
  number={1},
  pages={V25--V33},
  year={2009},
  publisher={Society of Exploration Geophysicists}
}

@article{anstey1966sectional,
  title={THE SECTIONAL AUTO-CORRELOGRAM AND THE SECTIONAL RETRO-CORRELOGRAM* PART I: THE SECTIONAL AUTO-CORRELOGRAM},
  author={Anstey, Nigel Allister},
  journal={Geophysical Prospecting},
  volume={14},
  number={4},
  pages={389--426},
  year={1966},
  publisher={Wiley Online Library}
}

@article{anstey1966part,
  title={PART II: THE SECTIONAL RETRO-CORRELOGRAM},
  author={Anstey, NA and Newman, P},
  journal={Geophysical Prospecting},
  volume={14},
  pages={411--426},
  year={1966},
  publisher={Wiley Online Library}
}

@article{verschuur1992adaptive,
  title={Adaptive surface-related multiple elimination},
  author={Verschuur, Dirk J and Berkhout, AJ and Wapenaar, CPA},
  journal={Geophysics},
  volume={57},
  number={9},
  pages={1166--1177},
  year={1992},
  publisher={Society of Exploration Geophysicists}
}

@article{chen2021fast,
  title={Fast high-resolution hyperbolic radon transform},
  author={Chen, Wei and Yang, Liuqing and Wang, Hang and Chen, Yangkang},
  journal={IEEE Transactions on Geoscience and Remote Sensing},
  volume={60},
  pages={1--10},
  year={2021},
  publisher={IEEE}
}

@inproceedings{ghosh2012multiple,
  title={Multiple suppression in the txp domain},
  author={Ghosh, Shaunak and Fomel, Sergey},
  booktitle={SEG International Exposition and Annual Meeting},
  pages={SEG--2012},
  year={2012},
  organization={SEG}
}

@inproceedings{herrmann2000aliased,
  title={De-aliased, high-resolution Radon transforms},
  author={Herrmann, Philippe and Mojesky, Todd and Magesan, Mag and Hugonnet, Pierre},
  booktitle={SEG International Exposition and Annual Meeting},
  pages={SEG--2000},
  year={2000},
  organization={SEG}
}

@inproceedings{hugonnet20093d,
  title={3{D} {H}igh resolution parabolic {R}adon filtering},
  author={Hugonnet, Pierre and Boelle, Jean-Luc and Mihoub, Majda and Herrmann, Philippe},
  booktitle={71st EAGE Conference and Exhibition incorporating SPE EUROPEC 2009},
  pages={cp--127},
  year={2009},
  organization={European Association of Geoscientists \& Engineers}
}

@incollection{sacchi1999fast,
  title={Fast high resolution parabolic Radon transform},
  author={Sacchi, Mauricio D and Porsani, Milton},
  booktitle={SEG Technical Program Expanded Abstracts 1999},
  pages={1477--1480},
  year={1999},
  publisher={Society of Exploration Geophysicists}
}

@article{foster1992suppression,
  title={Suppression of multiple reflections using the Radon transform},
  author={Foster, Douglas J and Mosher, Charles C},
  journal={Geophysics},
  volume={57},
  number={3},
  pages={386--395},
  year={1992},
  publisher={Society of Exploration Geophysicists}
}

@book{weglein2005multiple,
  title={Multiple attenuation},
  author={Weglein, Arthur B and Dragoset, William H},
  year={2005},
  publisher={Society of Exploration Geophysicists Tulsa, Oklahoma}
}

@incollection{hampson1986inverse,
  title={Inverse velocity stacking for multiple elimination},
  author={Hampson, Dan},
  booktitle={SEG Technical Program Expanded Abstracts 1986},
  pages={422--424},
  year={1986},
  publisher={Society of Exploration Geophysicists}
}

@ARTICLE{licmul2024,
  author={Li, Chao and Liu, Guochang and Chen, Xiaohong and Li, Zhichao and Fomel, Sergey and Chen, Yangkang},
  journal={IEEE Transactions on Geoscience and Remote Sensing}, 
  title={Joint {R}econstruction and {M}ultiple {A}ttenuation {U}sing {O}ne-{S}tep {R}andomized-{O}rder {D}amped {R}ank {R}eduction {M}ethod}, 
  year={2024},
  volume={62},
  number={},
  pages={1-11},
  doi={10.1109/TGRS.2024.3435560}
}

@article{lopezg2015,
    author = {Lopez, Gabriel A. and Verschuur, D. J.},
    title = {Closed-loop surface-related multiple elimination and its application to simultaneous data reconstruction},
    journal = {Geophysics},
    volume = {80},
    number = {6},
    pages = {V189-V199},
    year = {2015},
    doi = {10.1190/geo2015-0287.1}
}

@article{zhangd2021cl,
author = {Dong Zhang and Dirk Jacob (Eric) Verschuur},
title = {Closed-loop surface-related multiple estimation with full-wavefield migration-reconstructed near offsets for shallow water},
journal = {Geophysics},
volume = {86},
number = {5},
pages = {WC21-WC30},
year = {2021},
doi = {10.1190/geo2020-0879.1},
}

@article{saado2024,
    author = {Saad, Omar M. and Ravasi, Matteo and Alkhalifah, Tariq},
    title = {Noise attenuation in distributed acoustic sensing data using a guided unsupervised deep learning network},
    journal = {Geophysics},
    volume = {89},
    number = {6},
    pages = {V573-V587},
    year = {2024},
    doi = {10.1190/geo2024-0109.1}
}

@article{sergeylocal2007,
author = {Sergey Fomel},
title = {Local seismic attributes},
journal = {Geophysics},
volume = {72},
number = {3},
pages = {A29-A33},
year = {2007},
doi = {10.1190/1.2437573},

}

@article{neelamani2010,
    author = {Neelamani, Ramesh (Neelsh) and Baumstein, Anatoly and Ross, Warren S.},
    title = {Adaptive subtraction using complex-valued curvelet transforms},
    journal = {Geophysics},
    volume = {75},
    number = {4},
    pages = {V51-V60},
    year = {2010},
    month = {08},
    doi = {10.1190/1.3453425}
}

@article{Herrmann2008,
    author = {Herrmann, Felix J. and Wang, Deli and Verschuur, Dirk J. (Eric)},
    title = {Adaptive curvelet-domain primary-multiple separation},
    journal = {Geophysics},
    volume = {73},
    number = {3},
    pages = {A17-A21},
    year = {2008},
    month = {05},
    doi = {10.1190/1.2904986}
}

@article{zhangd2019,
    author = {Zhang, Dong and Verschuur, D. J. (Eric) and Qu, Shan and Chen, Yangkang},
    title = {Surface-related multiple leakage extraction using local primary-and-multiple orthogonalization},
    journal = {Geophysics},
    volume = {85},
    number = {1},
    pages = {V81-V97},
    year = {2019},
    month = {11},
    doi = {10.1190/geo2019-0465.1}
}

@article{zhangd2021f,
    author = {Zhang, Dong and Verschuur, D. J. (Eric) and Chen, Yangkang},
    title = {Fast local primary-and-multiple orthogonalization for surface-related multiple leakage estimation and extraction},
    journal = {Geophysics},
    volume = {86},
    number = {4},
    pages = {V353-V360},
    year = {2021},
    month = {07},
    doi = {10.1190/geo2020-0420.1}
}

@article{madiba2003,
    author = {Madiba, Gislain B. and McMechan, George A.},
    title = {Processing, inversion, and interpretation of a 2{D} seismic data set from the {N}orth {V}iking {G}raben, {N}orth {S}ea},
    journal = {Geophysics},
    volume = {68},
    number = {3},
    pages = {837-848},
    year = {2003},
    month = {01},
    doi = {10.1190/1.1581036}
}


\clearpage
\listoffigures

\AtEndDocument{
\clearpage
\begin{figure*}
\centering
\includegraphics[width=1\columnwidth]{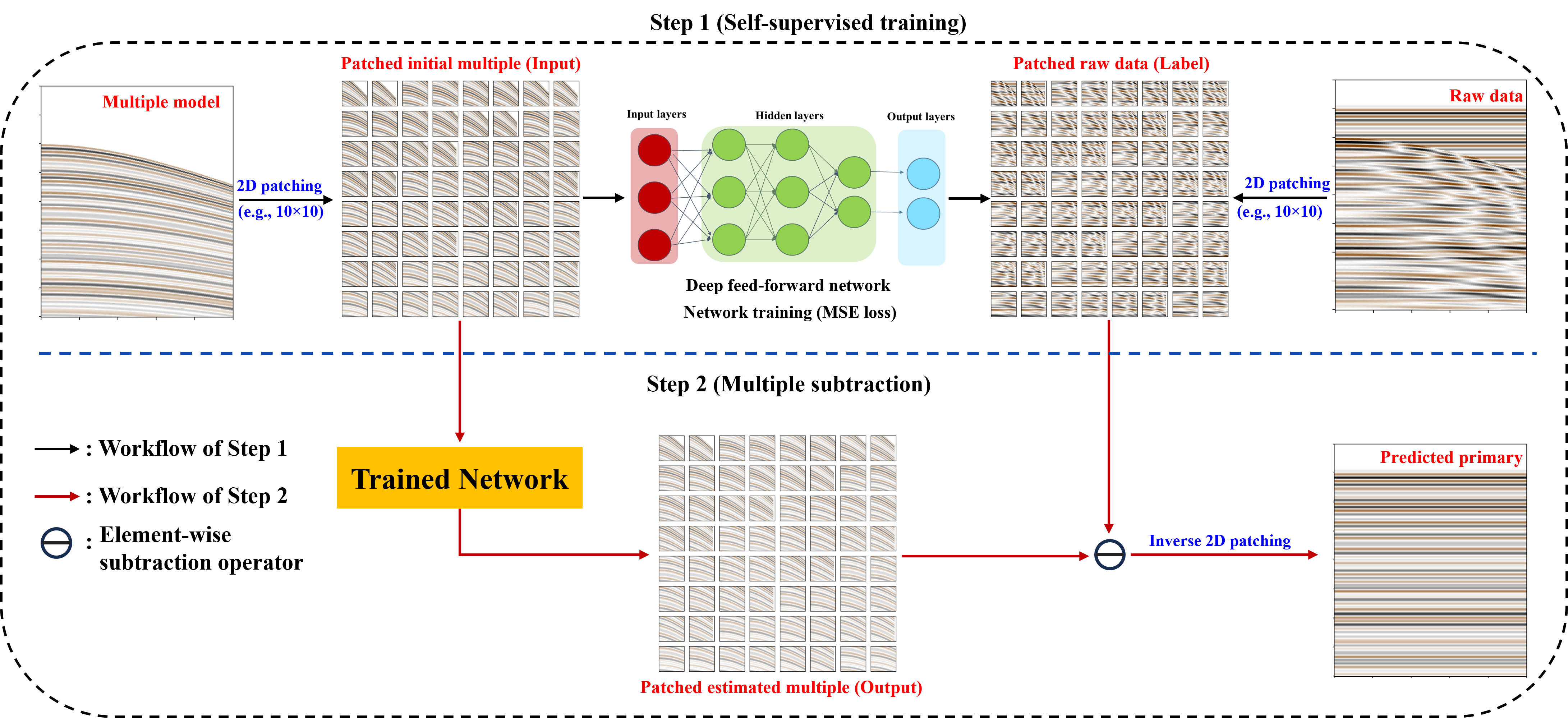}
\caption{Workflow of the proposed PDL method. In the PDL method, we employ a two-step scheme in a self-supervised manner, utilizing a patching scheme \cite{omar2021gp1} for local feature extraction and multiple predictions to enhance performance. }
\label{fig:pdlsketch}   
\end{figure*}

\clearpage
\begin{figure*}
\centering
\includegraphics[width=1\columnwidth]{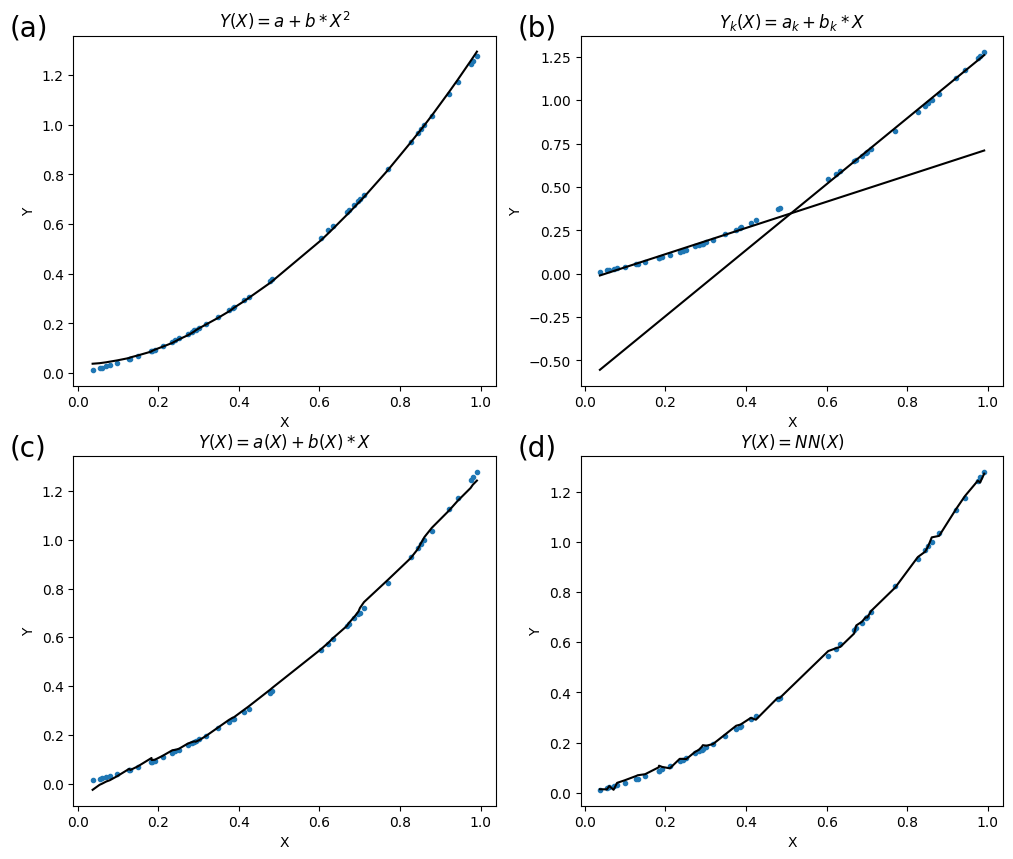}
\caption{A synthetic benchmark example comparing the various regression algorithms. (a) Scattered points follow a quadratic polynomial function $y=a+b*x^2$, and with additive random noise overlaid on a quadratic polynomial fitting curve. (b) The same scattered points overlaid by two linearly fitted lines on two parts of the samples. (c) Same scattered points overlaid by the non-stationary regression fitted curve \cite{fomel2009adaptive}. (d) Same scattered points overlaid by the neural network fitted curve.}
\label{fig:comp}   
\end{figure*} 

\clearpage
\begin{figure*}
\centering
\includegraphics[width=1\columnwidth]{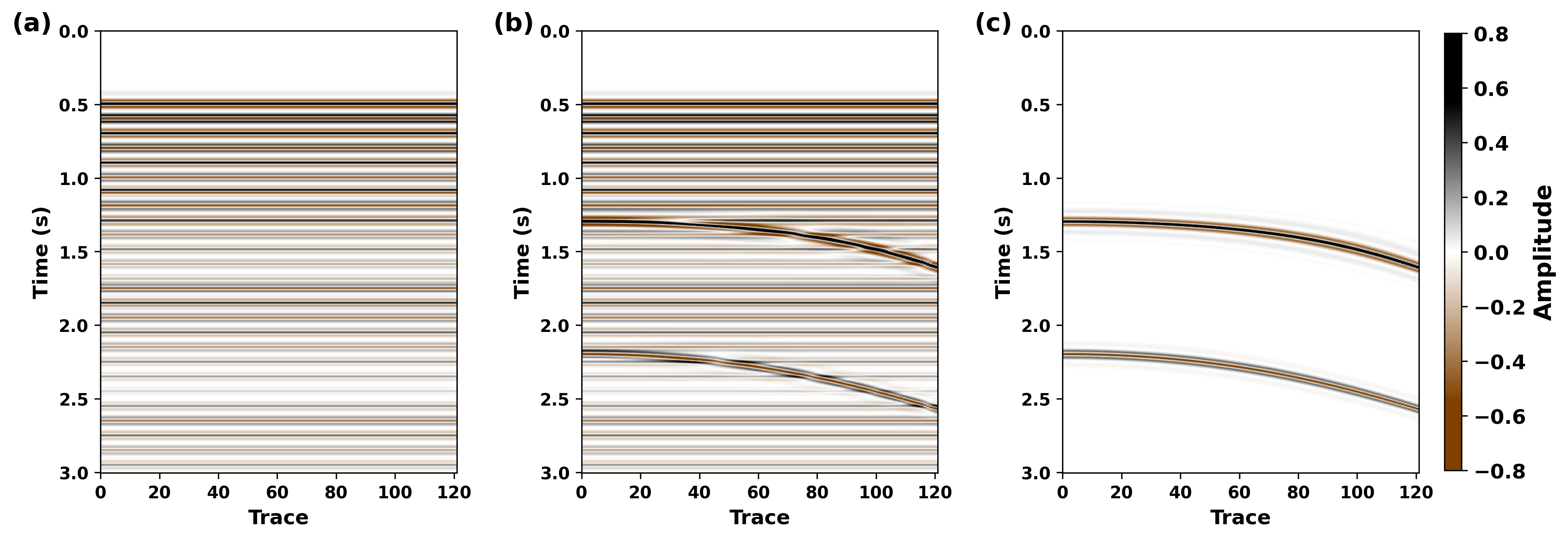}
\caption{The first synthetic data. (a) The real primaries. (b) The original data with multiples. (c) The estimated multiples with amplitude and phase differences.}
\label{fig:syndata1}   
\end{figure*} 

\clearpage
\begin{figure*}
\centering
\includegraphics[width=1.\columnwidth]{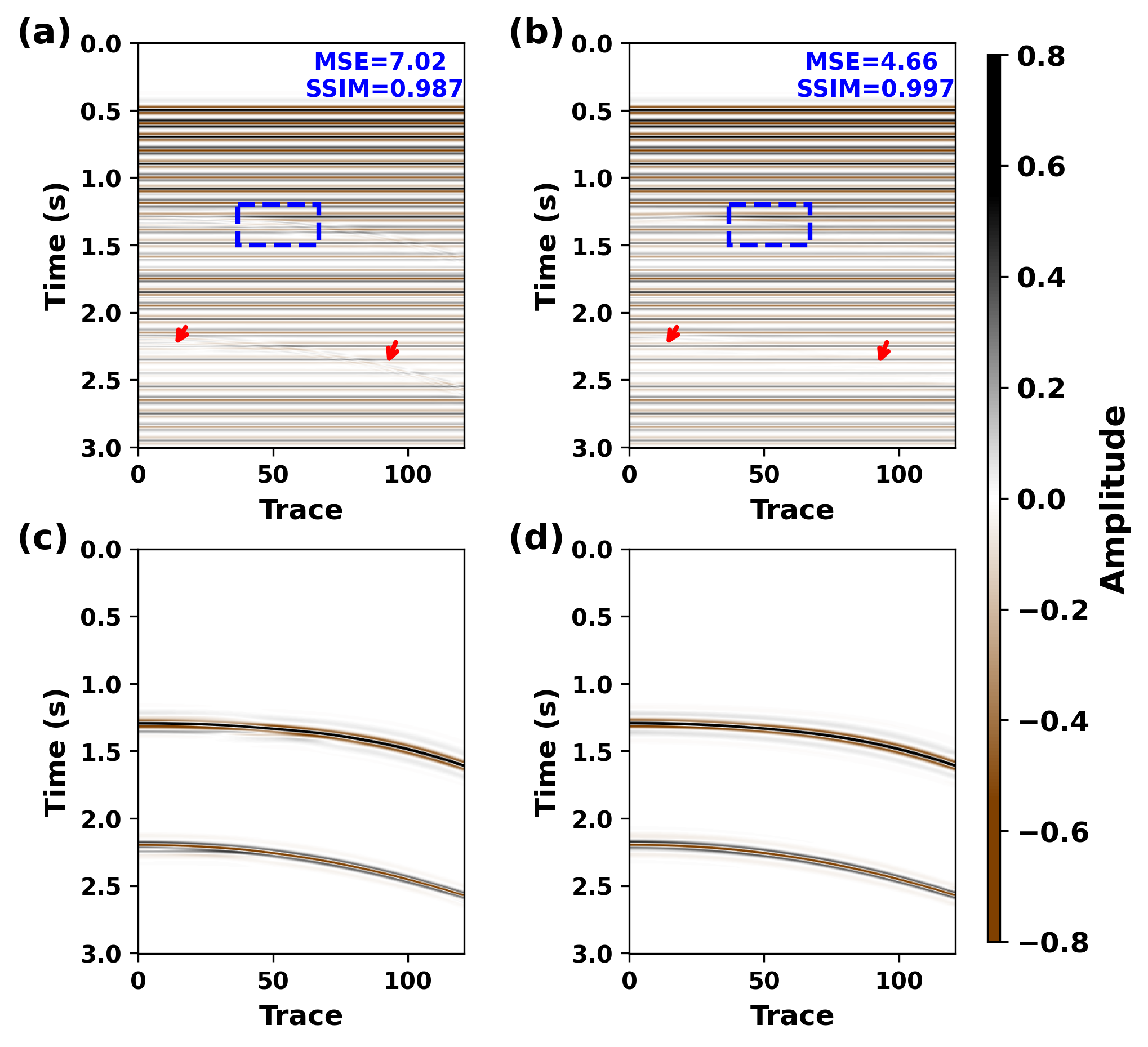}
\caption{(a) The estimated primary wave using the classic LSMF method. (b) The estimated primary wave using the PDL method. (c) The removed multiples wave using the classic LSMF method. (d) The removed multiples wave using the PDL method. The blue frame boxes are zoomed in Figure \ref{fig:zoomsyn1}.}
\label{fig:synpre1}
\end{figure*}

\clearpage
\begin{figure*}
\centering
\includegraphics[width=1.\columnwidth]{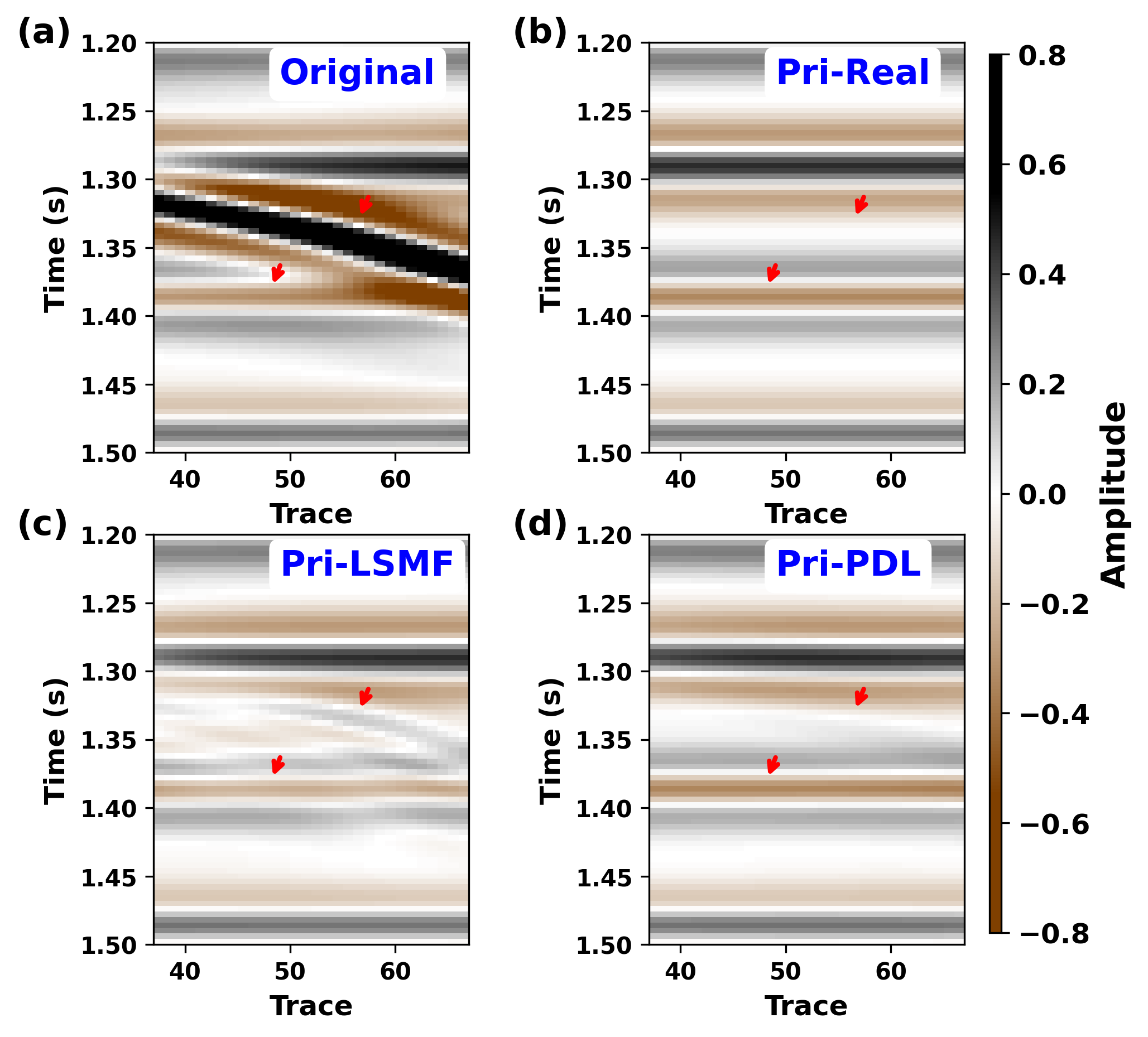}
\caption{(a) Zoomed section corresponding to Figure~\ref{fig:syndata1}b. (b) Zoomed section corresponding to Figure~\ref{fig:syndata1}a. (c) Zoomed section corresponding to Figure~\ref{fig:synpre1}a. (d) Zoomed section corresponding to Figure~\ref{fig:synpre1}b. }
\label{fig:zoomsyn1}
\end{figure*}

\clearpage
\begin{figure*}
\centering
\includegraphics[width=1\columnwidth]{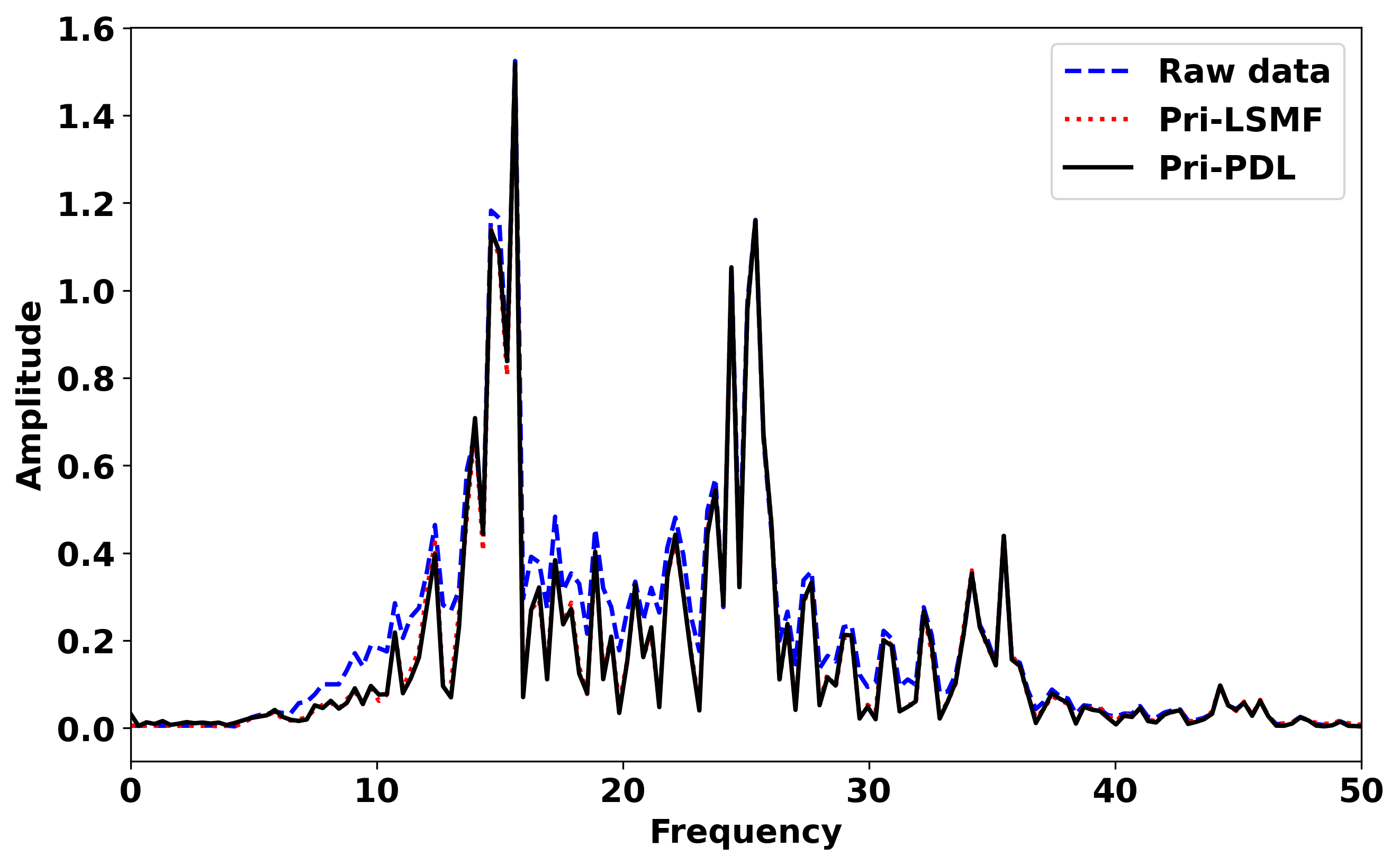}
\caption{Spectral comparison of raw data (Figure~\ref{fig:syndata1}b), estimated primaries using LSMF method (Figure~\ref{fig:synpre1}a), and estimated primaries using PDL method (Figure~\ref{fig:synpre1}b).}
\label{fig:synspe1}
\end{figure*}

\clearpage
\begin{figure*}
\centering
\includegraphics[width=1.\columnwidth]{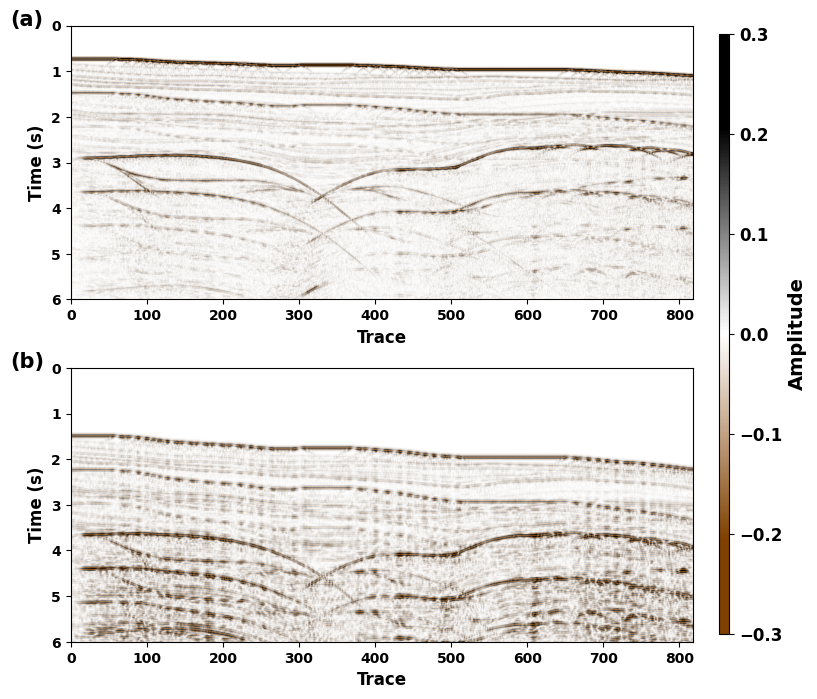}
\caption{The Pluto dataset (common offset domain). (a) The original data with multiples. (b) The estimated multiples with amplitude and phase differences.}
\label{fig:syndata2}
\end{figure*} 

\clearpage
\begin{figure*}
\centering
\includegraphics[width=1.\columnwidth]{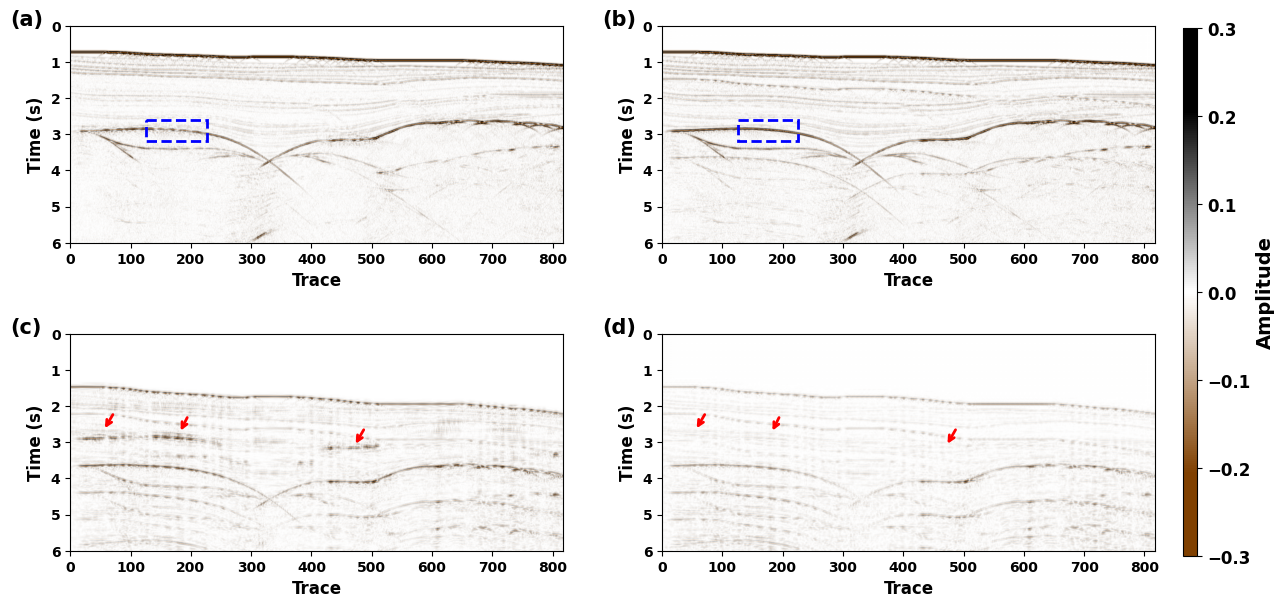}
\caption{(a) The estimated primary wave using the classic LSMF method. (b) The estimated primary wave using the PDL method. (c) The removed multiples using the classic LSMF method. (d) The removed multiples using the PDL method. The blue frame boxes are zoomed in Figure \ref{fig:zoomsyn2}.}
\label{fig:synpre2}
\end{figure*}

\clearpage
\begin{figure*}
\centering
\includegraphics[width=1.0\columnwidth]{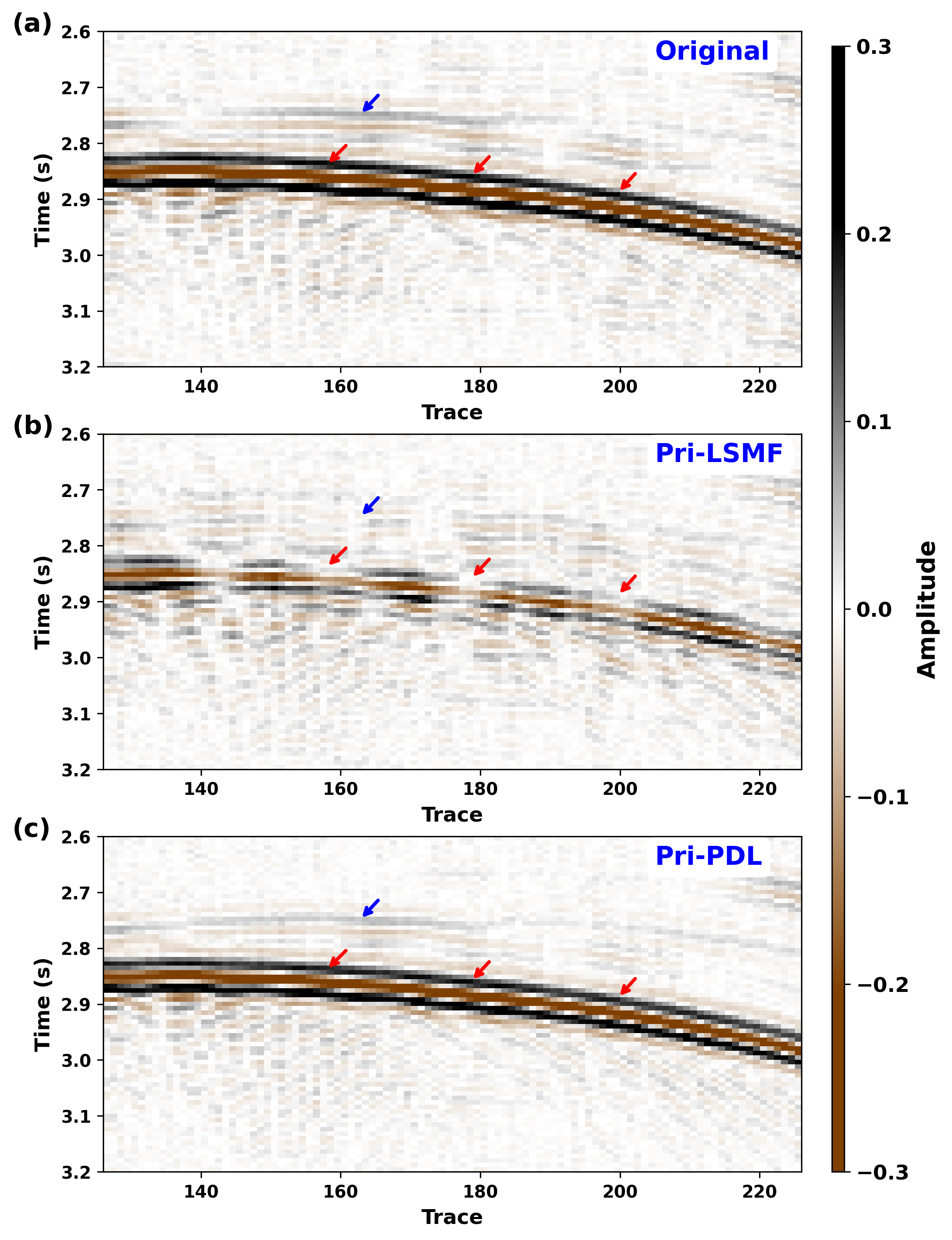}
\caption{(a) Zoomed section corresponding to Figure~\ref{fig:syndata2}a. (b) Zoomed section corresponding to Figure~\ref{fig:synpre2}a. (c) Zoomed section corresponding to Figure~\ref{fig:synpre2}b.}
\label{fig:zoomsyn2}
\end{figure*} 

\clearpage
\begin{figure*}
\centering
\includegraphics[width=1\columnwidth]{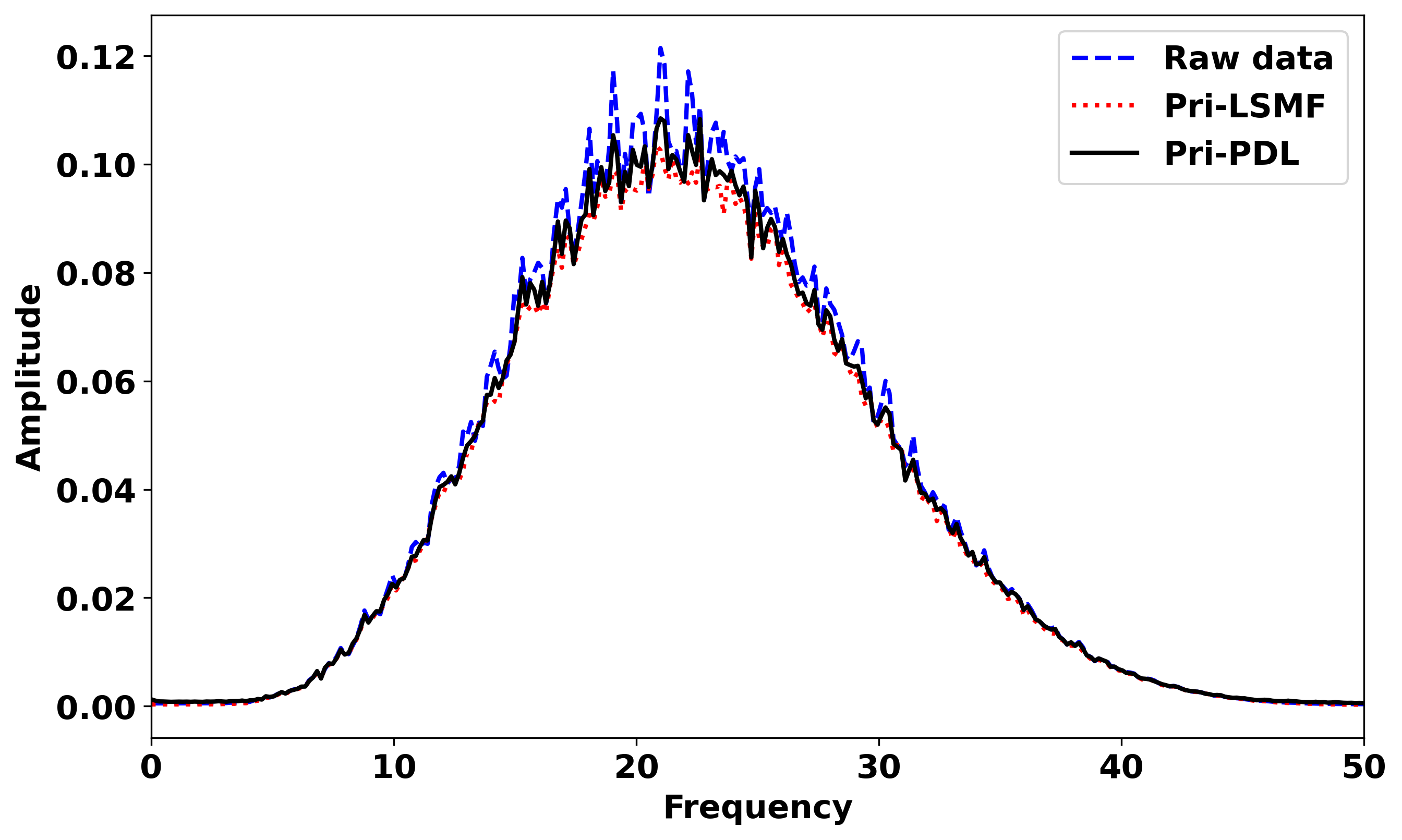}
\caption{Spectral comparison of raw data (Figure~\ref{fig:syndata2}a), estimated primaries using LSMF method (Figure~\ref{fig:synpre2}a), and estimated primaries using PDL method (Figure~\ref{fig:synpre2}b).}
\label{fig:synspe2}
\end{figure*}

\clearpage
\begin{figure*}
\centering
\includegraphics[width=1.0\columnwidth]{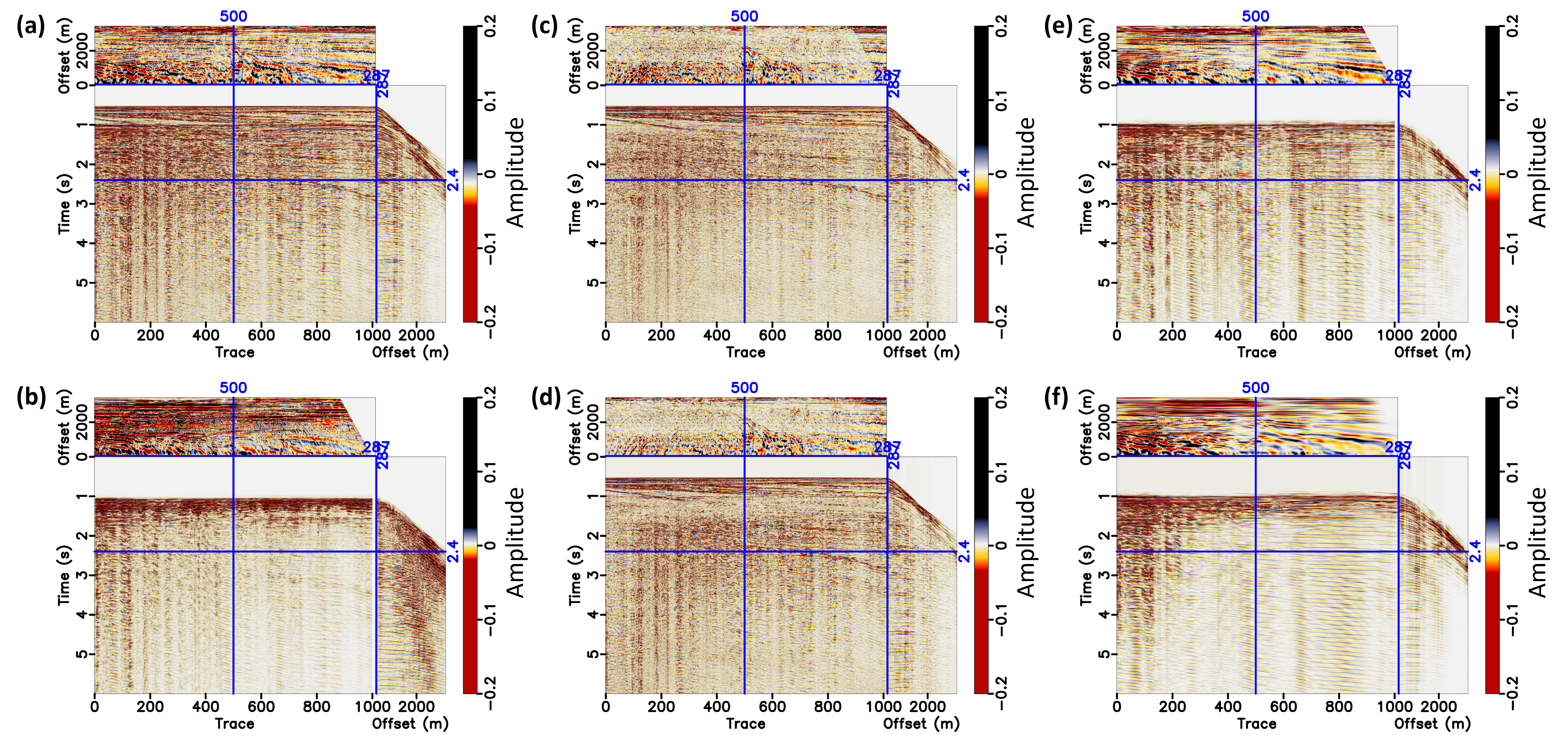}
\caption{Viking Graben field data example. (a) Shot gathers with multiples. (b) The estimated multiples using SRME. (c) The predicted primaries using the classic LSMF method. (d) The predicted primaries using the PDL method. (e) The removed multiples using the classic LSMF method. (f) The removed multiples using the PDL method.}
\label{fig:3dshotfld1}
\end{figure*}

\clearpage
\begin{figure*}
\centering
\includegraphics[width=1.0\columnwidth]{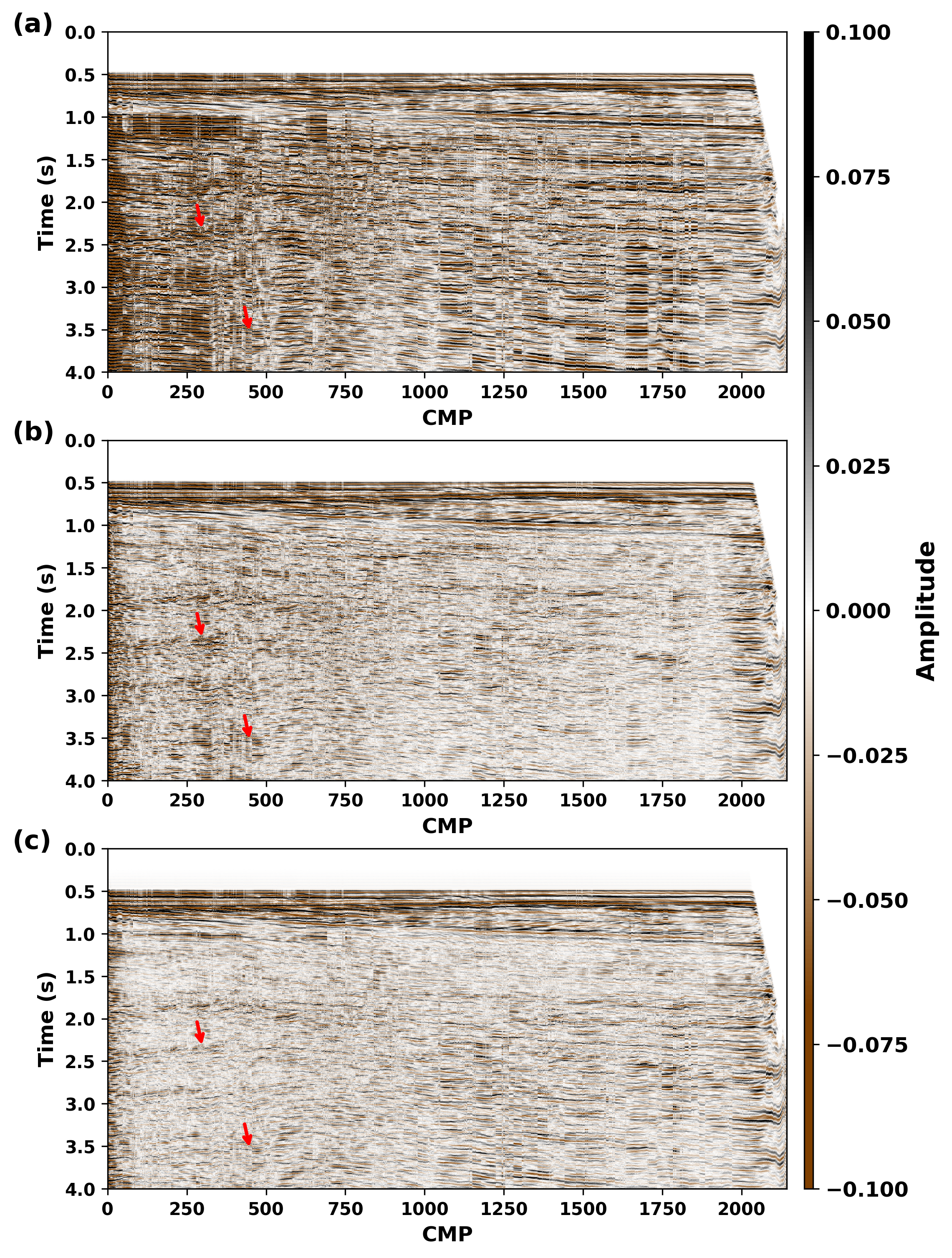}
\caption{\new{(a) Stacked section obtained from the raw data. (b) Stacked section obtained after multiple suppression using the LSMF method. (c) Stacked section obtained after multiple suppression using the proposed PDL method.}}
\label{fig:fldstacks}
\end{figure*}

\clearpage
\begin{figure*}
\centering
\includegraphics[width=1.0\columnwidth]{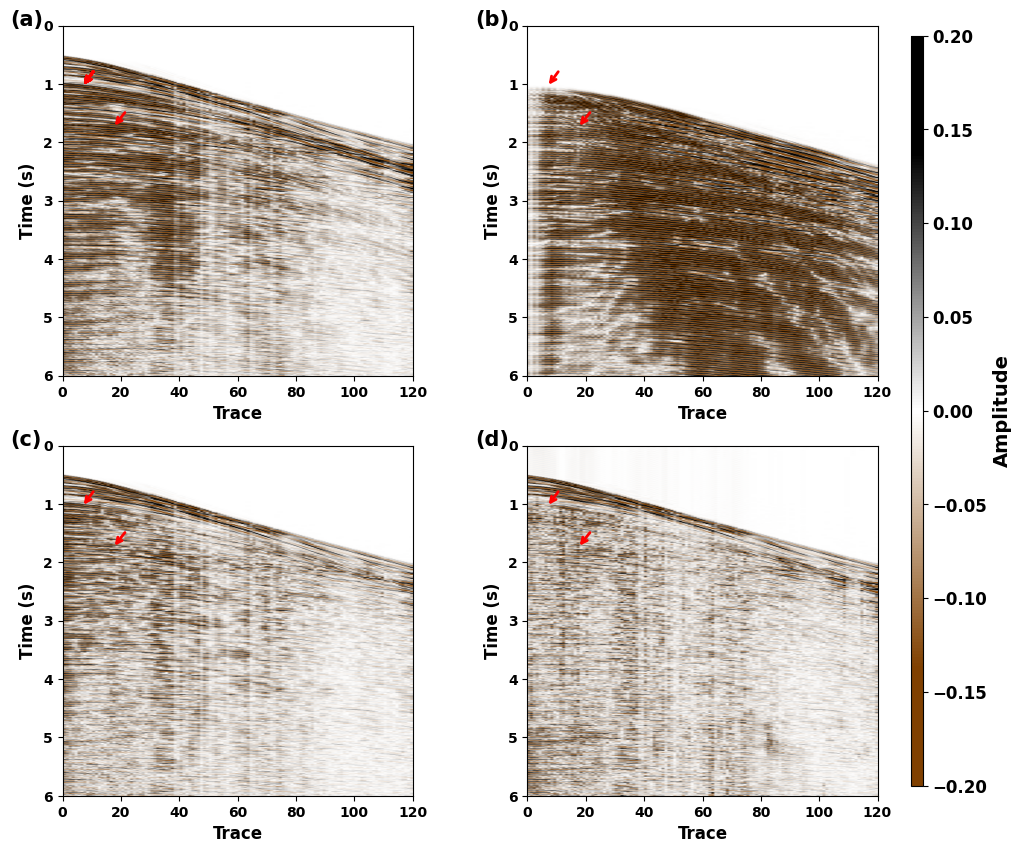}
\caption{(a) A single shot with multiples. (b) The estimated multiple model of a single shot using SRME. (c) The predicted primary wave of a single shot using the classic LSMF method. (d) The predicted primary wave of a single shot using the PDL method.}
\label{fig:shotfld1}
\end{figure*}

\clearpage
\begin{figure*}
\centering
\includegraphics[width=1\columnwidth]{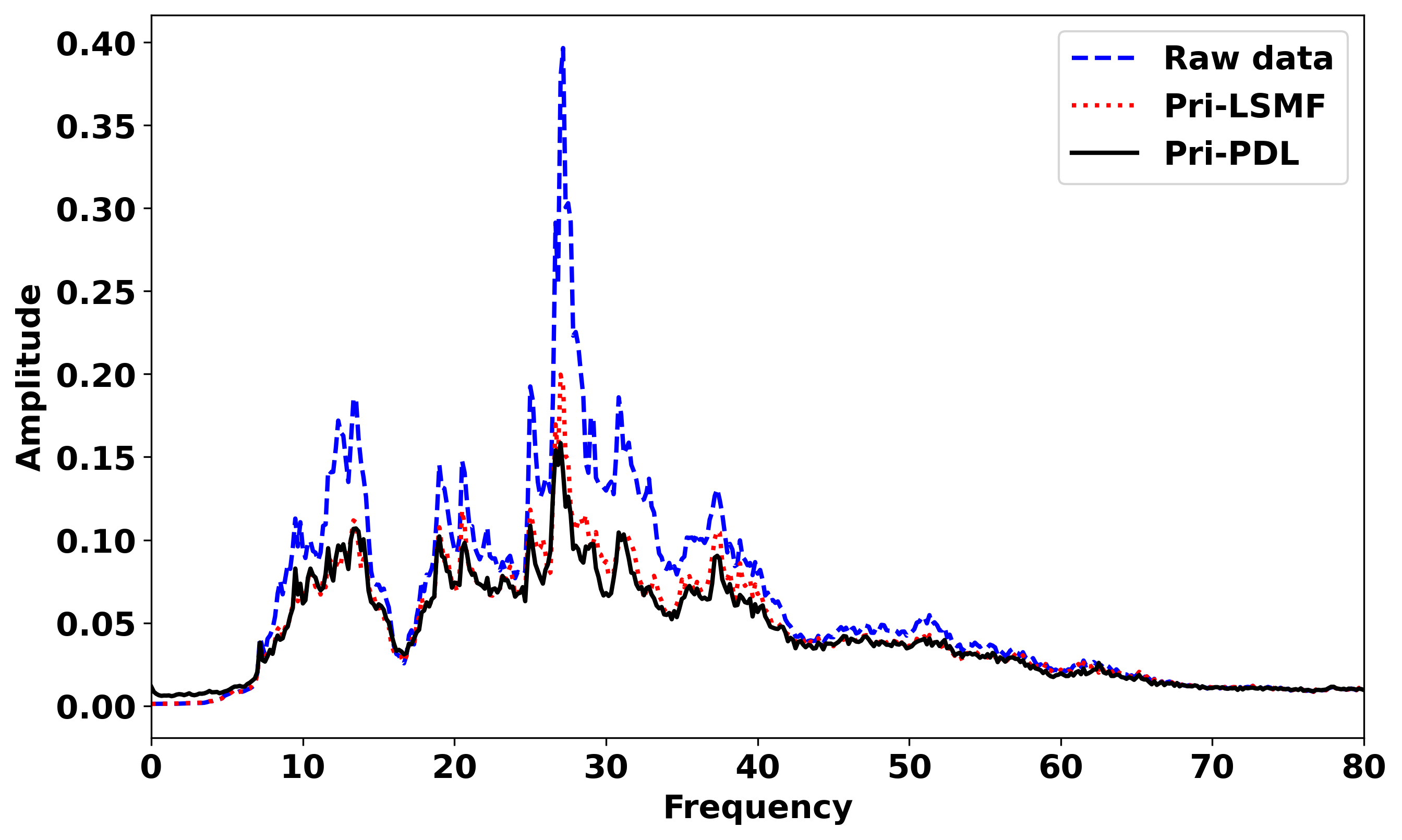}
\caption{Spectral comparison of raw data (Figure~\ref{fig:shotfld1}a), estimated primaries using LSMF method (Figure~\ref{fig:shotfld1}c), and estimated primaries using PDL method (Figure~\ref{fig:shotfld1}d). }
\label{fig:fldprespe1}
\end{figure*}

\clearpage
\begin{figure*}
\centering
\includegraphics[width=1.0\columnwidth]{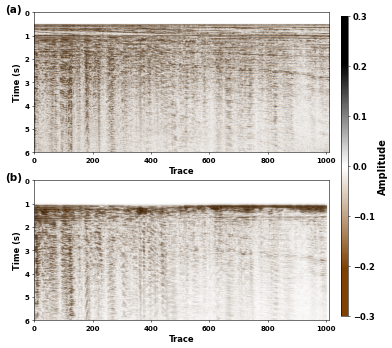}
\caption{(a) The original field data with multiples. (b) The estimated multiples of the field data using the SRME method.}
\label{fig:flddata1}
\end{figure*}

\clearpage
\begin{figure*}
\centering
\includegraphics[width=1.0\columnwidth]{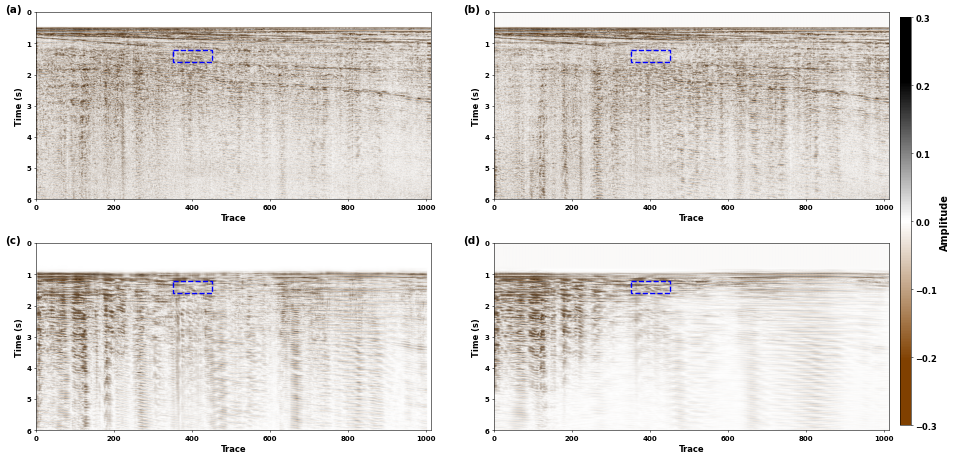}
\caption{(a) The estimated primary wave using the classic LSMF method. (b) The estimated primary wave using the PDL method. (c) The removed multiples using the classic LSMF method. (d) The removed multiples using the PDL method. The blue frame boxes are zoomed in Figure \ref{fig:zoomfld1}.}
\label{fig:fldpre1}
\end{figure*}

\clearpage
\begin{figure*}
\centering
\includegraphics[width=0.7\columnwidth]{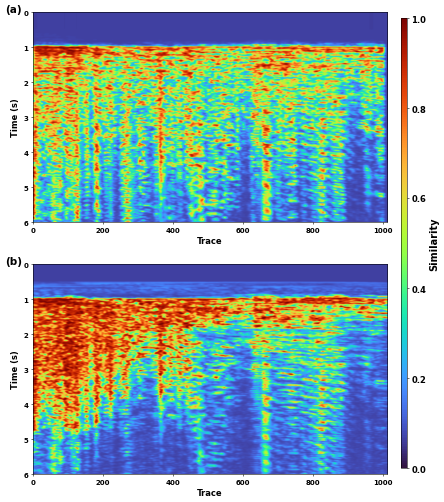}
\caption{(a) The local similarity coefficients between the raw field data and the removed multiple estimated using the LSMF method. (b) The local similarity coefficients between the raw field data and the removed multiple estimated using the PDL method. }
\label{fig:fldpresim1}
\end{figure*}

\clearpage
\begin{figure*}
\centering
\includegraphics[width=1.0\columnwidth]{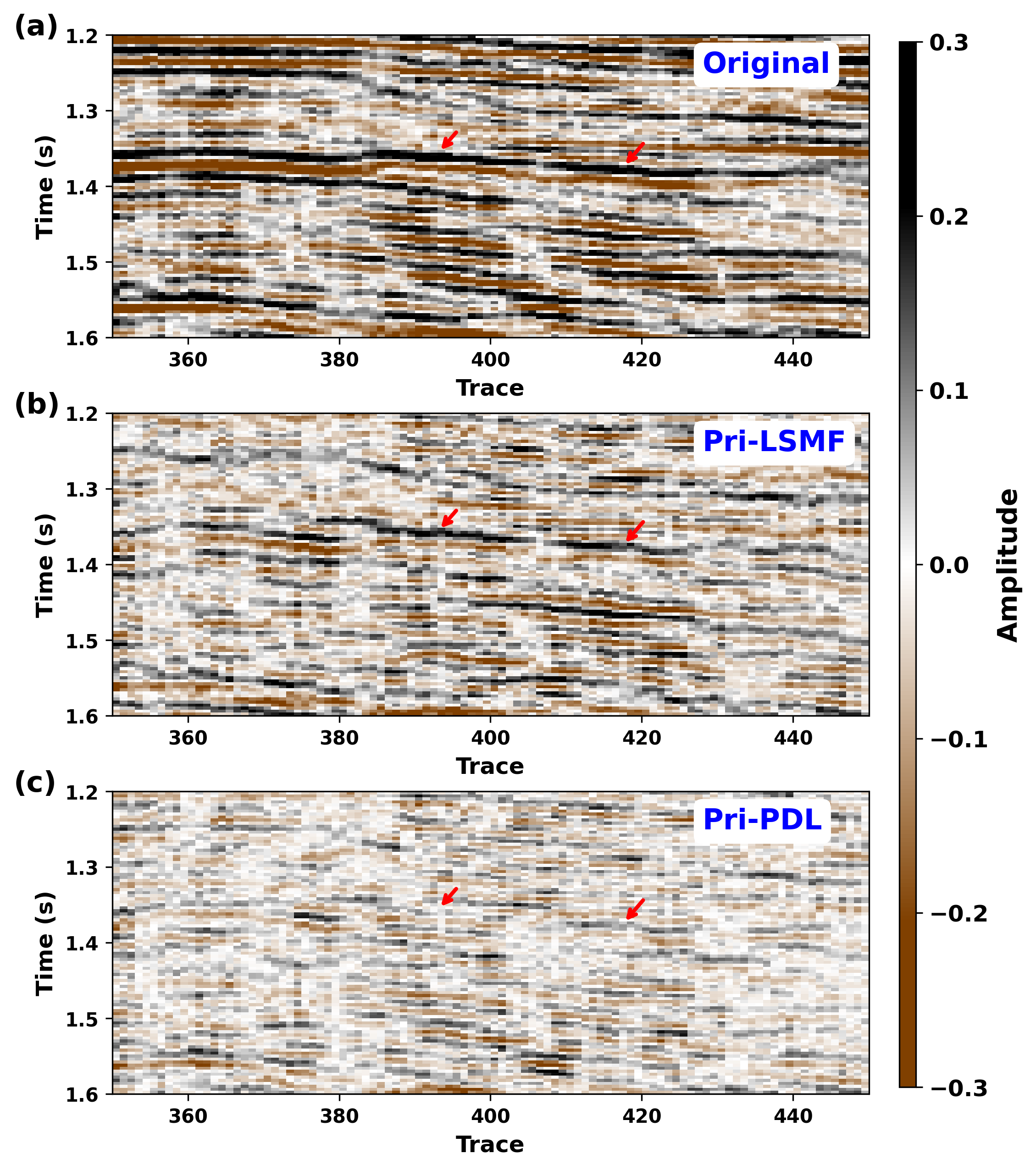}
\caption{(a) Zoomed section comparison corresponding to Figure~\ref{fig:flddata1}a. (b) Zoomed section comparison corresponding to Figure~\ref{fig:fldpre1}a. (c) Zoomed section comparison corresponding to Figure~\ref{fig:fldpre1}b.}
\label{fig:zoomfld1}
\end{figure*}

\clearpage
\begin{figure*}
\centering
\includegraphics[width=0.7\columnwidth]{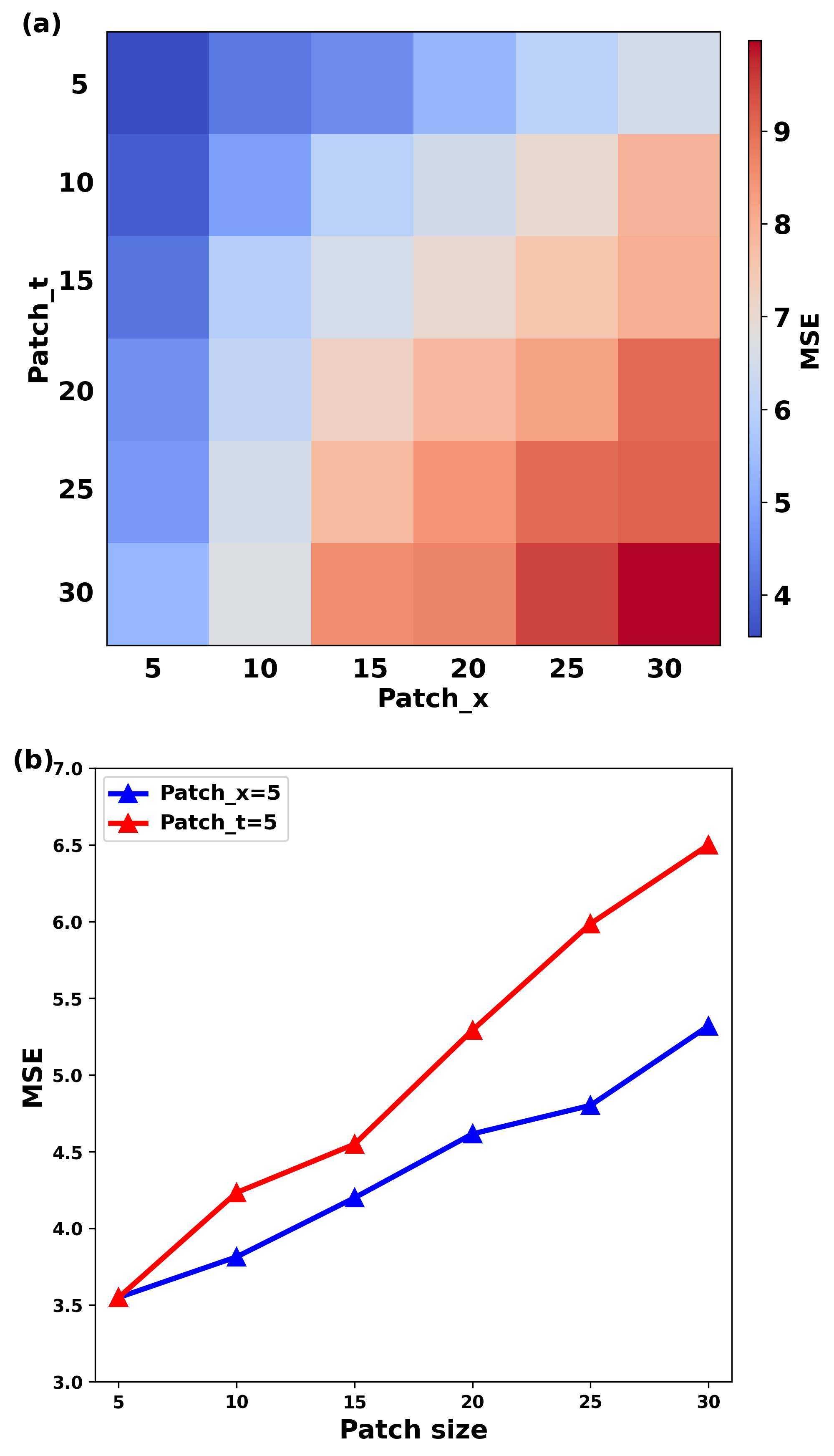}
\caption{(a) MSE variation with different patch sizes. (b) MSE curves with fixed Patch\_x or Patch\_t.}
\label{fig:patchsize}
\end{figure*}

\clearpage
\begin{figure*}
\centering
\includegraphics[width=1\columnwidth]{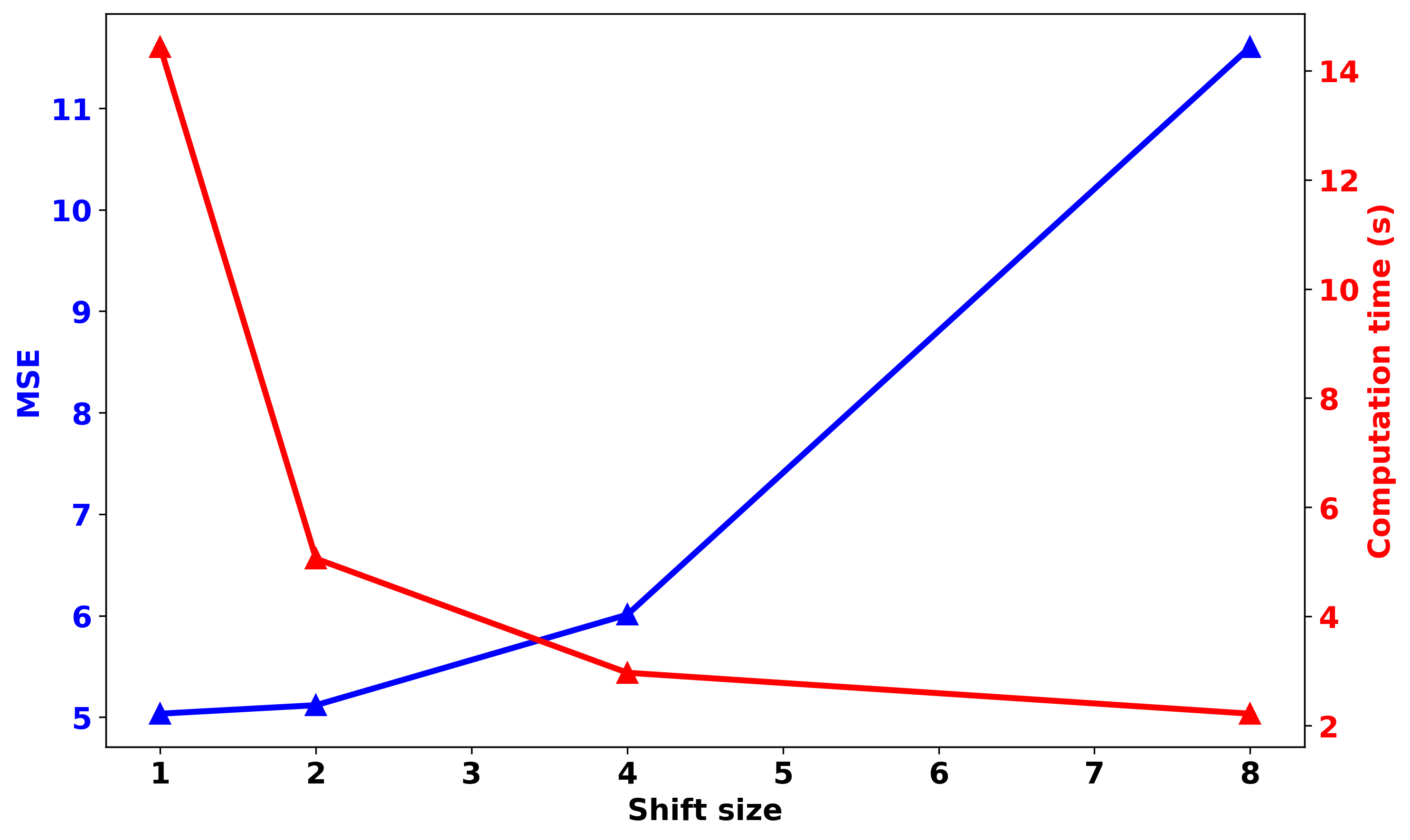}
\caption{MSE and computation time variation with different shift sizes.}
\label{fig:shiftsize}
\end{figure*}

\clearpage
\begin{figure*}
\centering
\includegraphics[width=1\columnwidth]{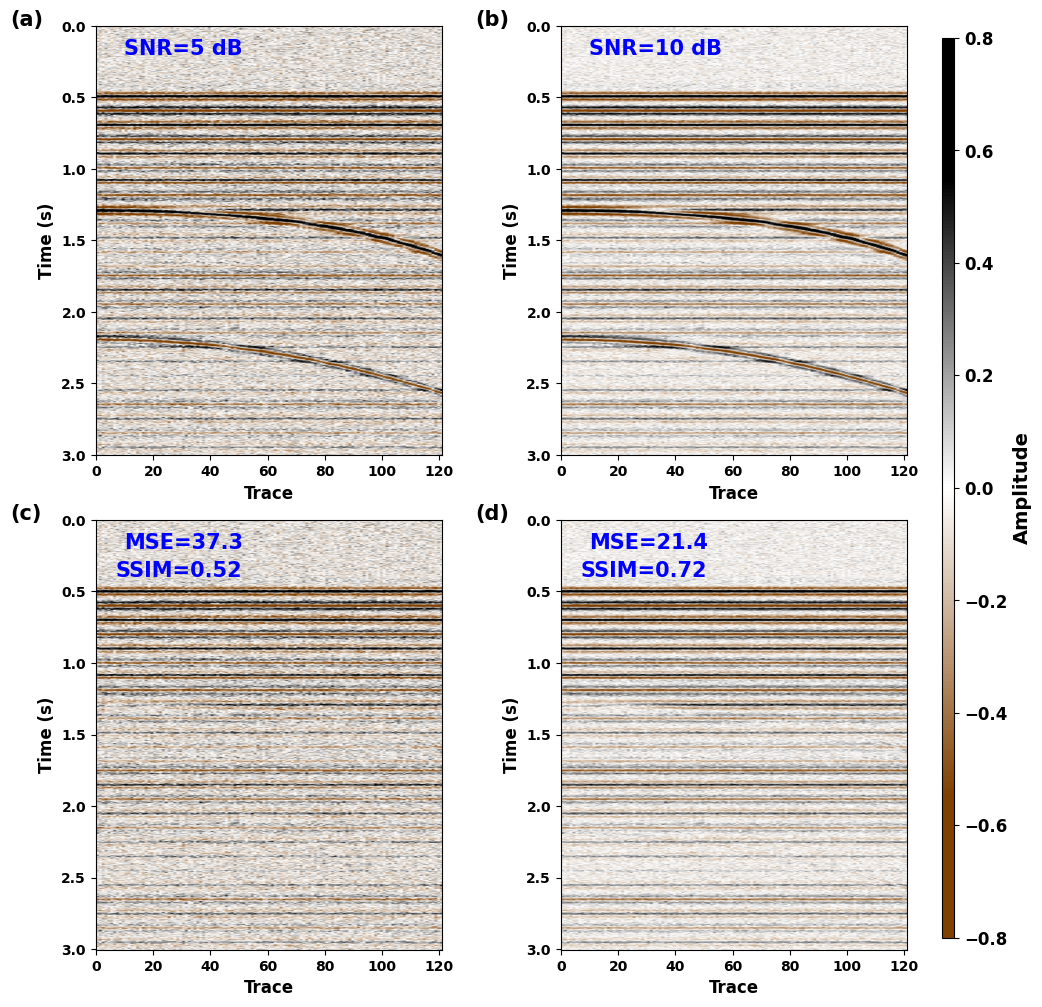}
\caption{(a) Noisy seismic data with an SNR of 5 dB. (b) Noisy seismic data with an SNR of 10 dB. (c) Predicted primaries based on the noisy data in Figure~\ref{fig:snrs}a. (d) Predicted primaries based on the noisy data in Figure~\ref{fig:snrs}b.}
\label{fig:snrs}
\end{figure*}

}
\end{document}